\documentclass[sigconf,nonacm]{acmart}
\usepackage{algorithm}
\usepackage{algorithmic}
\usepackage{booktabs,makecell,siunitx}
\usepackage{array}
\usepackage{fancyvrb}
\usepackage{float}
\usepackage{framed}
\usepackage{graphicx}
\usepackage{subcaption}
\usepackage{color}

\newcommand{\System}{\textsc{Bonsai}}

\usepackage{marginnote}

\usepackage{url}
\AtBeginDocument{%
  }

\copyrightyear{2026}
\acmYear{2026}
\setcopyright{cc}
\setcctype{by}
\acmConference[CHI '26]{Proceedings of the 2026 CHI Conference on Human Factors in Computing Systems}{April 13--17, 2026}{Barcelona, Spain}
\acmBooktitle{Proceedings of the 2026 CHI Conference on Human Factors in Computing Systems (CHI '26), April 13--17, 2026, Barcelona, Spain}
\acmDOI{10.1145/3772318.3791855}
\acmISBN{979-8-4007-2278-3/2026/04}

\begin{document}

\title[Bonsai: Intentional and Personalized Social Media Feeds]{
\texorpdfstring{\includegraphics[width=\baselineskip]{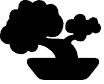}}{}{\textsc{BONSAI}}: Intentional and Personalized Social Media Feeds
}

\begin{teaserfigure}
\centering
  \includegraphics[scale=.975]{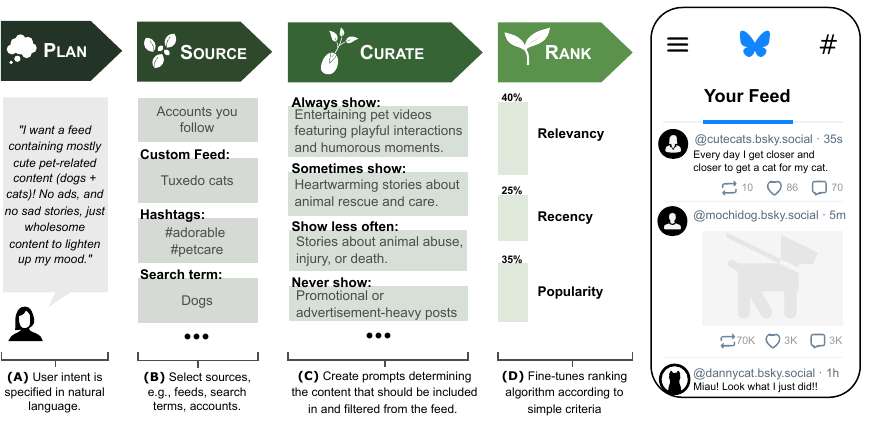}
  \caption{\System{} empowers users to create \textit{intentional and personalized social media feeds} on Bluesky via natural language-based prompts.
  Feed creation is simple yet customizable and transparent, users
  (A) specify in broad strokes what they want to obtain from their feed;
  (B) select content sources that will power the feed;
  (C) specify a series of prompts determining what should or should not be on the feed;
  (D) choose the priorities of the ranking algorithm.
  This process creates a feed tailored to the users' stated preferences, which can be revised to their liking at any time.
}
  \label{fig:teaser}
\end{teaserfigure}

\author{Omar El Malki}
\email{omar.elmalki@epfl.ch}
\affiliation{%
  \institution{EPFL}
  \city{Lausanne}
  \state{Vaud}
  \country{Switzerland}
}
\authornote{Work done while at Princeton.}

\author{Marianne Aubin Le Qu\'er\'e}
\email{marianne.alq@princeton.edu}
\affiliation{%
  \institution{Princeton University}
  \city{Princeton}
  \state{New Jersey}
  \country{USA}
}

\author{Andrés Monroy-Hernández}
\email{andresmh@cs.princeton.edu}
\affiliation{%
  \institution{Princeton University}
  \city{Princeton}
  \state{New Jersey}
  \country{USA}
}

\author{Manoel Horta Ribeiro}
\affiliation{%
  \institution{Princeton University}
  \city{Princeton}
  \state{New Jersey}
  \country{USA}
}
\email{manoel@cs.princeton.edu}

\renewcommand{\shortauthors}{El Malki et al.}

\begin{abstract}
Social media feeds use predictive models to maximize engagement, often misaligning how people consume content with how they wish to.
We introduce \System{}, a system that enables people to build \textit{personalized} and \textit{intentional} feeds.
\System{} implements a platform-agnostic framework comprising Planning, Sourcing, Curating, and Ranking modules. 
This framework allows users to express their intent in natural language and exert fine-grained control over a procedurally transparent feed creation process.
We evaluated the system with 15 Bluesky users in a two-phase, multi-week study. 
We find that participants successfully used our system to discover new content, filter out irrelevant or toxic posts, and disentangle engagement from intent, but curating intentional feeds required more effort than they are used to.
Simultaneously, users sought system transparency mechanisms to effectively use (and trust) intentional, personalized feeds.
Overall, our work highlights intentional feedbuilding as a viable path beyond engagement-based optimization.
\end{abstract}


%


\maketitle

\vspace{10mm}
\begin{framed}
\begin{center}
\textbf{    Work published at CHI 2026. Please cite accordingly.}
\end{center}
\end{framed}
\newpage
\section{Introduction}

At the heart of modern social media feeds are predictive models designed to maximize user engagement~\cite{metaOurApproachFacebook, zhao2019recommending, liu2022monolith}. 
These algorithms analyze vast amounts of behavioral data to forecast which posts a user is most likely to view, like, share, or comment on~\cite{narayanan2023understanding}.
Platforms like Bluesky, X, Facebook, Instagram, and TikTok rely on this approach to rank and personalize content, effectively shaping each user’s feed into a unique, constantly adapting experience. 
This formula aligns tightly with advertising-based business models that monetize attention and impressions, giving platforms strong incentives to optimize for engagement~\cite{kim2017social}.
It also has clear short-term benefits: engagement-driven ranking can make apps feel enjoyable and relevant~\cite{reddits}, and users spend more time in feeds curated by these predictive systems than in chronological ones~\cite{guessHowSocialMedia2023}. 

At the same time, many users experience a disconnect between how they use social media in ``the spur of the moment'' and how they wish they used it in retrospect~\cite{narayanan2023understanding}. 
In behavioral science \textit{lingo}~\cite{kahneman2011thinking}, this dichotomy reflects a tension between System 1 (fast, intuitive, and reactive) and System 2 (slower, deliberate, and reflective). 
While scrolling through a feed, users are often driven by immediate rewards like novelty or emotional stimulation. 
Afterwards, however, users may feel their usage did not align with deeper goals, such as staying informed or connecting meaningfully to others~\cite{tranModelingEngagementDisengagementCycle2019, kleinberg2024challenge, milli2025engagement}.
This gap between use and intent contributes to compulsive habits~\cite{tran2019modeling}, problematic or addictive use~\cite{marinoComprehensiveMetaanalysisProblematic2018, sapaczAreWeAddicted2016}, and even decisions to deactivate platforms altogether~\cite{allcott2020welfare, tromholt2016facebook}.

But how can users increase their agency while using social media?
Past work in HCI and CSCW has explored a range of approaches to help people feel more in control of their feeds and their time: from topic-based interfaces that reduce information overload~\cite{bernsteinEddiInteractiveTopicbased2010} to introducing friction to disrupt mindless scrolling~\cite{hinikerMyTimeDesigningEvaluating2016, gruningDirectingSmartphoneUse2023}.
On decentralized social media platforms in particular, communities have built customization tools, such as user-authored feeds on Bluesky (e.g., SkyFeed), that let people specify rules, keywords, and filters to shape what appears in their timelines.
These systems show that some users are motivated to take a more active role in configuring their feeds, and that even relatively simple customization can make platforms feel more aligned with their goals.

Yet existing approaches face important limitations.
Time management and friction tools can reduce overuse, but they are blunt instruments, preventing use altogether or restricting access to entire apps, rather than helping users transform \emph{how} they use those apps.
Rule-based customization tools can support fine-grained control, but, in practice, require substantial effort and technical understanding. Users must translate high-level goals (e.g., ``see more thoughtful posts from climate policy experts'') into low-level rules, lists, and filters, and must continually adjust these configurations as their interests and the platform evolve.
As a result, many customization tools remain niche, and engagement-optimized feeds continue to dominate the default user experience, powered by the vast amounts of implicit feedback signals that platforms have~\cite{covington2016deep}.

In that context, language models (LMs) may change the rules of the game. 
Unlike traditional machine learning systems or manual interventions, LMs can follow natural language instructions and adapt to different contexts without task-specific retraining, large labeled datasets, or constant user maintenance~\cite{kojima2022large,wang2025end}. 
These capabilities may enable easy-to-use, intent-driven personalization. Instead of relying on engagement signals inferred from past behavior, recommender systems and feed algorithms can now incorporate direct user input about what they want to see. 
They could do so at a level of specificity (e.g., tone, sources, trade-offs between relevance and serendipity) that would be cumbersome to encode as rules.
Further, such feeds can offer more granular, legible control: prompts and configuration settings directly specify which content enters the feed and can be inspected, discussed, and iteratively refined.

To investigate the feasibility and characteristics of intentional, personalized feeds, we adopt an exploratory,
systems-based approach. Broadly, we ask: \emph{How do LM-mediated feed builders reshape how people author, experience, and
sustain personalized social media feeds?}
We investigate this overarching direction with \System{}, a system designed to increase user agency by helping them create personalized feeds that respond to intentions communicated in transparent, natural language prompts.
We implemented \System{} for Bluesky, an open and decentralized social media platform, as illustrated in Fig.~\ref{fig:teaser}, and conducted a two-phased interview study ($n=15$) to understand how users design and experience intentional feeds. Given that \System{} explores a new design space, we evaluate it ``in a holistic fashion on the basis of what [the system] makes possible,''  focusing on ``how compelling, how richly painted, and how informed is [the system's] vision.'' \cite{wobbrock2012seven}.

\System{} was guided by three design goals: \textit{intent fidelity}, ensuring that explicitly stated user goals govern feed generation; \textit{ease of use}, lowering the cost of preference articulation so users can flexibly adjust feeds without specialized expertise; 
and \textit{procedural transparency}, making the sourcing, filtering, and ranking of content visible and understandable. 
Together, these goals aim to shift personalization away from engagement optimization toward feeds intentionally authored by users.
To operationalize these goals, \System{} implements a platform-agnostic four-stage framework.
First, in the \textit{Planning} stage, an LM agent parses the user’s natural language description into candidate sources and inclusion/exclusion criteria; 
Second, in the \textit{Sourcing} stage, we fetch posts from feeds, lists, accounts, hashtags, or search queries from platform APIs, LMs, and previously scraped databases;
Third, in the \textit{Curation} stage, an LM agent scores candidate posts against user-stated preferences.
Finally, in the \textit{Ranking} stage, posts are ordered based on user-specified trade-offs among relevance, recency, and popularity. The result is an editable configuration in which users iteratively co-author the algorithm that shapes their social media experience.

Our findings suggest that intentional, personalized feeds help participants discover relevant content, focus on desired information, and avoid overload, ultimately strengthening their sense of agency.
The LM-mediated nature of~\System{} enabled participants to express nuanced preferences, convey social media intentions through precise post specifications, and allowed novice users to onboard easily to custom feeds.
At the same time, these feeds demand greater effort to construct and heighten user expectations that the resulting content will faithfully reflect their goals; while some users find excitement and fulfillment in building intentional social feeds, others struggle with the cognitive load associated with the task.
Participant feedback highlights the promise of hybrid approaches that combine intentional and engagement-based feeds.

Through this work, we aim to concretely demonstrate the feasibility of intentional, personalized social media through natural language. 
Overall, we contribute:

\begin{enumerate}

\item \System{}, a system that empowers intentional and personalized feedbuilding through explicit natural language input. To the best of our knowledge, \System{} is the first end-user-facing system using language models to enable people to author and control social media feeds in natural language.

\item Results from a multi-week field study with 15 Bluesky users
that characterize how people actually author, experience, and sustain
LM-mediated intentional feeds with \System{}

\item Three design implications for future systems that move beyond \System{}: Transparency makes granular controls actionable; Tightening the feedback loop; Doubling down on natural language (see \S\ref{designimp}).

\end{enumerate}

\section{Background and Related Work}

We build on research on the limits of engagement-optimized feeds, exploring systems that expand user control over technology use, and studies of transparency and user understanding of algorithms.

\subsection{Aligning Feeds with Explicit User Goals}
\label{rw:intent}

Most social media platforms rank content based on \textit{predicted} engagement. 
That is, they train classifiers to estimate the likelihood that a user will interact with a post (e.g., clicking, watching, or commenting) and use these predictions as the main signals for ranking~\cite{cunningham2024we}. 
This paradigm has proved highly effective at sustaining attention: engagement-optimized feeds consistently outperform chronological or deterministic baselines. 
For instance, Twitter’s long-term chronological holdback group received about 38\% fewer impressions per day as of November 2021~\cite{bandy2023exposure}. In 2020, Meta ran experiments with semi-chronological feeds, which led to substantial declines in time spent---20\% for Facebook, 10\% for Instagram~\cite{guessHowSocialMedia2023}.
Even Reddit, once known for its relatively simple “Hot” feed, ultimately adopted engagement-based personalization~\cite{reddits}.

Yet maximizing engagement often conflicts with users’ explicit goals. 
Experimental and observational work shows that engagement-driven feeds can promote addictive and regrettable use~\cite{tranModelingEngagementDisengagementCycle2019, cho2021reflect}, exacerbate compulsive habits~\cite{tran2019modeling}, and surface uncivil or emotionally charged content that users later report not wanting to see~\cite{guessHowSocialMedia2023, milli2025engagement}. 
Deactivation studies find that participants report improved well-being after leaving social media altogether~\cite{tromholt2016facebook, allcott2020welfare, allcott2025effect}, which may not be all that surprising since previous work suggests that around one in four people worldwide exhibit signs of problematic or addictive use~\cite{cheng2021prevalence}. 
These patterns have fueled broad concerns about the societal harms of engagement-optimized feeds~\cite{kleinberg2024challenge, harris2016technology}.

In response, platforms have begun to incorporate \textit{non-engagement signals} into ranking. 
One approach is to train classifiers on content ``quality'' metrics, such as YouTube’s downranking of sensationalist or tabloid-like videos~\cite{youtube_recommendation_system_2021}. 
Another is to use survey-based signals, where users rate specific items (e.g., ``Is this worth your time'') or platform-wide experiences (e.g., ``Are you receiving too many clickbait-y videos today?''), and predictive models generalize these responses at scale~\cite{cunningham2024we}. 
Researchers have pushed this logic further by calling for \textit{normative values} as first-class ranking objectives~\cite{stray2024building, bernstein2023embedding}. 
For example, \citet{jia2024embedding} introduced a ``democratic attitude'' signal into ranking and showed that down-ranking anti-democratic content reduced partisan animosity without degrading user experience, an approach recently extended into a broader pluralistic library of values~\cite{kolluri2025alexandria}.

In parallel, a growing body of work in Human--AI Interaction shows that users do not simply consume algorithmic outputs, but actively try to steer and collaborate with AI systems. Interactive and ``teachable'' systems let people correct or refine model behavior through examples, labels, or natural-language feedback, and find that users are most willing to provide such input when it clearly influences outcomes and remains lightweight~\cite{amershi2014power,ramos2020interactive,feng2024mapping}. In this view, aligning feeds with explicit goals is part of a broader shift toward mixed-initiative systems, where humans articulate high-level intentions while the AI infers low-level details and proposes refinements over time~\cite{horvitz1999principles}.

Our work builds on these initiatives by moving past implicit proxies altogether. 
Rather than relying on platform-wide quality metrics or abstract normative signals, we enable users to directly articulate their intentions in natural language and build feeds aligned with their stated goals. 
This approach resonates with the broader vision of ``teachable'' feed systems~\cite{feng2024mapping}, in which users without Machine Learning expertise iteratively shape algorithmic behavior through explicit feedback~\cite{ramos2020interactive}.

\begin{table*}[t]
\small
    \centering
    \caption{Description of systems that support more intentional technology use and feed control.}
    \label{tab:feed-systems-comparison}
    \begin{tabular}{p{1.5cm}p{9.3cm}p{4.2cm}}
        \toprule
        \textbf{System} & \textbf{Mechanism} & \textbf{Level of abstraction for users} \\
        \midrule
        \textsc{MyTime}~\cite{hinikerMyTimeDesigningEvaluating2016} 
        & User-defined schedules to reduce time spent in “poor-use” apps while preserving “good-use” ones. 
        & App-level time blocks. \\ \midrule
        
        \textsc{NUGU}~\cite{koNUGUGroupbasedIntervention2015} 
        & Group-based accountability and shared goals to curb the overuse of smartphones. 
        & High-level usage goals and social check-ins. App-level time blocks. \  \\ \midrule
        
        \textsc{OneSec}~\cite{gruningDirectingSmartphoneUse2023} 
        & Inserts intentional delays and breathing prompts before opening selected apps to interrupt habitual use. 
        & Simple app-level switches.  \\ \midrule
        
        \textsc{Eddi}~\cite{bernsteinEddiInteractiveTopicbased2010} 
        & Clusters feed items into topical groups and lets users browse and filter by topic. 
        & Topic-level browsing and filtering; lightweight manual exploration. \\ \midrule
        
        \textsc{Cura}~\cite{he2023cura} 
        & Predicts which posts a designated curator would endorse from community upvotes and generates curated feeds. 
        & Community-level preferences inferred by the model. \\ \midrule
        
        \textsc{Locus}~\cite{davisSupportingTeensIntentional2023} 
        & Prompts users to choose a session purpose (e.g., learning vs.\ socializing), and reflect upon usage
        & High-level session goals and reflections. \\ \midrule
        
        \textsc{Skyfeed} and \textsc{Graze}\footnotemark[1] 
        & Visual builders and filters allow users to define custom feed rules over follows, keywords, and interactions. 
        & Low-level rules, filters, and regex. \\ \midrule
                
        \textbf{\System{} }
        & High-level goals described in natural language. Language models map these goals to editable, language-first configurations for sourcing, curating, and ranking content.
        & High-level natural-language prompts. \\
        \bottomrule
    \end{tabular}
\Description{The table compares eight systems that support more intentional technology use and feed control along three columns: System, Mechanism, and Level of abstraction for users. It lists MyTime, NUGU, OneSec, Eddi, Cura, Locus, Skyfeed \& Graze, and Bonsai. Earlier systems focus on scheduling or friction (e.g., MyTime, OneSec), group accountability (NUGU), topic or curator-based feeds (Eddi, Cura), or session goals (Locus), with user interaction at the app, topic, or rule level. Skyfeed and Graze require users to specify low-level rules and filters, whereas Bonsai is distinguished as using high-level natural-language prompts that are mapped by language models into configurations for sourcing, curating, and ranking content.
}
\end{table*}

\subsection{User Agency and Control in Feeds}
\label{rw:control}
As smartphones and computers provide access to seemingly endless streams of content, many individuals report feeling less able to focus and less ``in control'' of their own time~\cite{sapaczAreWeAddicted2016,marinoComprehensiveMetaanalysisProblematic2018,alter2018irresistible}. 
To regain control, users often take deliberate breaks from technology, whether by abstaining from social media during specific periods~\cite{schoenebeckGivingTwitterLent2014}, giving up their smartphones altogether~\cite{lee2014supporting}, or deleting their accounts~\cite{portwood-stacerMediaRefusalConspicuous2013}. 
Amidst these practices, and calls to study the \textit{non-use} of technology~\cite{satchellUserUseNonuse2009}, HCI researchers have developed systems that support self-regulation of digital engagement~\cite{lyngsSelfControlCyberspaceApplying2019}. 

Early interventions primarily focused on \emph{when} and \emph{how much} technology was used. 
For example, \textsc{MyTime} provided lightweight scheduling to reduce time spent in “poor-use” apps while preserving “good-use” ones~\cite{hinikerMyTimeDesigningEvaluating2016}, 
\textsc{NUGU} leveraged group accountability to curb overuse~\cite{koNUGUGroupbasedIntervention2015}, 
and \textsc{OneSec} inserted intentional delays to disrupt habitual app opening~\cite{gruningDirectingSmartphoneUse2023}. 
These systems were effective at mitigating compulsive and `undesired' use. For instance \textsc{MyTime} reduced usage on apps that participants felt were a poor use of time by 21\% while keeping the usage of `useful' apps unchanged. Yet, note that these systems intervene at the temporal level of use, rather than shaping the \emph{content} that fills feeds. 

A parallel line of work has explored richer, user-driven controls within social media feeds themselves. 
\textsc{Eddi} enabled interactive topic browsing by clustering feed items into trending topics; users found browsing to be more efficient, enjoyable, and less overwhelming than chronological interfaces, often noting they could ``find things […] faster'' and avoid irrelevant content~\cite{bernsteinEddiInteractiveTopicbased2010}; 
\textsc{Cura}, on the other hand, uses transformer-based modeling to predict which posts a designated curator would endorse based on community upvotes; evaluations demonstrated that its curated feeds accurately reflected curator preferences, adapting noticeably when curators changed, and halving antisocial behavior without extra moderation effort~\cite{he2023cura}. \textsc{Locus} scaffolds intentional entry points such as choosing a session purpose (e.g., "learning" vs. "socializing"), particularly helping younger users select content aligned with their intentions~\cite{davisSupportingTeensIntentional2023}. 
On Bluesky, we note two third-party tools: \textsc{Skyfeed} and \textsc{Graze}.
These tools let users author custom feed logic through visual rule builders, regular expressions, and composable filters,\footnote{E.g., \url{https://skyfeed.app}, \url{https://www.graze.social}.} exemplifying the promise of ``algorithmic choice,'' while still requiring users to think in terms of low-level rules.

With the increasing capabilities of language models, a growing body of work has explored \textit{natural language input} as a mechanism for communicating preferences to algorithms~\cite{zhou2024language,gao2021advances}. 
\citet{wang2025end} compared three strategies for end-user content curation: example labeling, rule writing, and prompting language models.
They found that natural language prompts enabled users to rapidly bootstrap personalized classifiers with strong performance. 
This insight aligns with a broader tradition of conversational recommendation, where preference elicitation through naturalistic questions often outperforms rigid item-attribute queries~\cite{kostric2021soliciting} and
recent work showing that language models can be integrated directly into recommendation pipelines~\cite{sanner2023large}.
However, despite this technical promise, LLM-powered recommenders have not yet seen widespread deployment in production feeds, in part due to cost, latency, and safety challenges at scale. Major platforms are only beginning to experiment with these ideas in their core products~\cite{threads_2025,instagram_2025,musk2025tweet}

This prior work, summarized in Table~\ref{tab:feed-systems-comparison}, motivates \System{} to shift from regulating when people use technology to shaping \emph{what} they see. Unlike prior systems that either operate at the temporal level (e.g., scheduling and friction tools), 
or require users to hand-craft low-level rules and filters, \System{} elicits user goals directly through natural language and makes each stage of feed construction explicit and editable.

\subsection{Transparency and Algorithmic Feeds}
\label{rw:transparency}

Users often misunderstand how algorithmic feeds work, and these misperceptions shape both trust and behavior~\cite{rader2015understanding,eslami2016first}. 
For instance, almost a decade ago, \citet{eslami2016first} found that many Facebook users were unaware their News Feed was curated at all.
They exposed participants to a tool called  \textsc{FeedVis}, which compared people's raw and algorithmically curated feeds.
This probe revealed and reshaped users’ \textit{folk theories} of curation, i.e., their working understandings of how the system operates~\cite{gelman2011concepts}.
Folk theories are consequential: they drive user behavior, which in turn influences algorithmic outcomes~\cite{devito2018people}. For example, when platform changes violate expectations, users often resist based on their ingrained folk theories, such as in the \#RIPTwitter backlash when users objected to proposed algorithmic changes~\cite{devito2017algorithms}.
Other work also shows how folk theories structure how marginalized users perceive algorithmic suppression of their identities and mobilize resistance in platforms like TikTok~\cite{karizat2021algorithmic}.

Algorithmic transparency can be understood as a way to approximate individuals' folk theories of how technology (here, algorithms) actually works~\cite{gelman2011concepts,devito2017algorithms,karizat2021algorithmic}.
As algorithms become more consequential in our lives, a vast body of work in (human-centered) explainable AI has sought to understand \textit{how} to close this gap, and bring transparency to algorithmic systems~\cite{ehsan2020human,wang2019designing,liao2021human}.
Past work toward this broad direction suggests that transparency is not monolithic but multi-faceted: it can involve documenting models and data~\cite{mitchell2019model}, publishing evaluations~\cite{liangholistic}, providing explanations~\cite{ribeiro2016should}, or communicating uncertainty~\cite{bhatt2021uncertainty}.

Further, past work shows that the effectiveness of transparency depends heavily on the audience: what helps developers or regulators may not support end-users, and vice versa~\cite{liao2023ai}. Still, across stakeholders, transparency is consistently framed as key to enabling meaningful control---whether through debugging, contestation, or audit---and as a mechanism for calibrating user trust and reliance~\cite{jacovi2021formalizing,liao2021human}. However, this literature has also suggested that \textit{transparency can backfire}, for instance by overwhelming users, fostering misplaced trust, or encouraging people to game the system~\cite{springer2020progressive,kizilcec2016much}. In that context, scholars caution that transparency without actionability risks creating false senses of agency~\cite{ananny2018seeing,smith2020no}, and that effective design requires progressive disclosure and tailoring to stakeholders’ contexts and capacities~\cite{springer2020progressive,liao2023ai}.

A parallel, broader line of work in Human--AI Interaction examines how transparency shapes ongoing collaboration between people and AI systems. Rather than treating explanations as purely informational, this research shows that users appropriate them to calibrate trust, adjust strategies, and figure out how to ``help'' the system behave as desired~\cite{kim2023help,liao2023ai}. For example, in an image-recognition setting, \citet{kim2023help} find that participants use model explanations to improve their own inputs and to diagnose when the AI has misunderstood a case, not just to verify predictions. More broadly, human-centered explainability work argues that transparency is most beneficial when it is tightly coupled to concrete levers for action, so that users can translate understanding into effective interventions~\cite{ehsan2020human,liao2021human}.

Transparency in feed algorithms has focused on helping users make sense of a black-box model after the fact, e.g., surfacing factors correlated with recommendations or disclosing which signals influence ranking~\cite{tintarev2015explaining}. Other efforts have emphasized exposing the assumptions and data inputs used to personalize feeds~\cite{rader2015understanding,rader2018explanations}. In contrast, \System{} offers a distinct approach: rather than retrofitting explanations onto engagement-based algorithms, it makes the feed construction pipeline itself transparent. This shifts the focus away from what \citet{ananny2018seeing} critique as ``black-box'' ideals of transparency~ toward a model in which users gain clarity by actively authoring the algorithmic process that shapes their feeds.  

\begin{figure}[t]
    \centering
    \includegraphics[width=\linewidth]{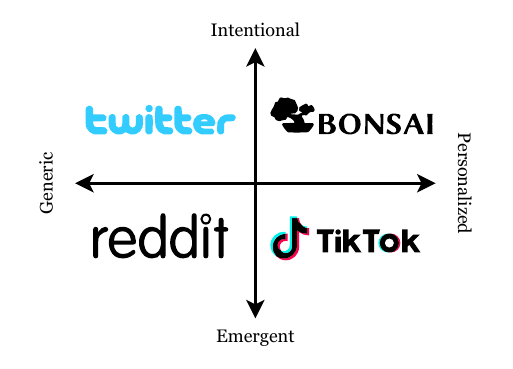}
    \caption{Positioning of different social media feeds along two dimensions: intentional vs. emergent and generic vs. personalized. 
    These axes are introduced in this paper as an analytic framing and do not reflect platform categories.
    Twitter's old chronological feed is generic and intentional;
    Reddit’s ``Hot'' feed is generic and emergent; TikTok’s ``For You'' feed is personalized and emergent. 
    \System{} appears in the remaining quadrant as a \emph{system for constructing feeds}, not a platform.
    }

    \Description{This figure presents a two-dimensional framework for categorizing social media feeds. The vertical axis contrasts intentional feeds, where users explicitly state their goals, with emergent feeds, where the system infers preferences from behavior. The horizontal axis contrasts generic feeds, which apply simple, uniform ranking rules, with personalized feeds, which tailor content to individual users. In this framework, Twitter’s old chronological feed is placed in the intentional–generic quadrant because users explicitly chose accounts to follow, but ranking was non-personalized and straightforward. Reddit’s ``Hot'' feed falls in the emergent–generic quadrant, since it relies on aggregate community votes and recency rather than individual user intent. TikTok's ``For You''' feed is positioned in the emergent–personalized quadrant, as it heavily personalizes recommendations based on inferred engagement patterns. Finally, Bonsai is placed in the intentional–personalized quadrant, representing feeds explicitly authored by users in natural language but also tailored to their preferences through algorithmic curation.
}
\label{fig:framework}    
\vspace{-1mm}
\end{figure}

\section{\includegraphics[width=\baselineskip]{imgs/bonsaipdf.pdf}\System{}: Design and System}

\System{} explores the design space for social media feeds to be both \textit{intentional} and \textit{personalized}.
Before detailing the specific design goals, we distinguish social media feeds according to two dimensions: intentional vs. emergent and generic vs. personalized.

\vspace{1mm}
\noindent
\textbf{Intentional vs. Emergent Feeds.}
A feed is \textit{intentional} when people's experience is shaped by the goals they explicitly articulate, e.g., by selecting topics or accounts to feature.  
Conversely, \textit{emergent} feeds derive from patterns of everyday use, with the system inferring intent from engagement and other in-platform behaviors.

\vspace{1mm}
\noindent
\textbf{Generic vs. Personalized Feeds.} 
A feed is \textit{generic} when it follows the same rule-set for all users, such as ranking posts by recency or aggregate popularity.  
By contrast, a \textit{personalized} feed adapts to the individual, ordering content according to user-specific features like past interactions, profile data, or inferred preferences.  \vspace{2mm}

We illustrate this taxonomy with three prominent social media feeds in Fig.~\ref{fig:framework}. 
TikTok's \textit{``For You Feed''} is \textit{emergent} and \textit{personalized}; recommendations are primarily driven by predicted user engagement, and your engagement with content strongly shapes the content you are shown~\cite{vombatkere2024tiktok}.
Reddit's \textit{``Hot Feed''} is \textit{emergent} and \textit{generic}; the feed is created with a simple formula that considers recency, but also engagement from others, including comment rate and upvote score~\cite{reddits}.
Old Twitter's \textit{``Chronological Feed''} is \textit{intentional} and \textit{generic}; the feed draws from sources that users explicitly choose to follow, and ranks their posts based on general criteria (recency) that do not consider engagement signals~\cite{devito2017algorithms}. 
In contrast, \System{} is in the \textit{intentional} and \textit{personalized} quadrant, enabling feeds tailored to stated intentions while still adapting to individual preferences. 

\subsection{Design Goals}

The central purpose of \System{} is to realize a feed that is both \emph{intentional} and \emph{personalized}. 
We operationalize this purpose by pursuing three design goals detailed below: \emph{intent fidelity}, \emph{ease of use}, and \emph{procedural transparency}.

\vspace{-1mm}
\subsubsection{{Intent fidelity}.}
\textit{Explicitly stated intentions govern the sourcing, filtering, and ranking of content in the feed.}
Users should have an interface that allows them to specify directly what they want---and do not want---to see.  
This goal responds to prior work (\S\ref{rw:intent}) showing that optimizing for \textit{revealed} preferences (e.g., clicks, dwell time) often drives regrettable or addictive use ~\cite{tranModelingEngagementDisengagementCycle2019,cho2021reflect,milli2025engagement}.  
This lets users isolate the personalization signal to their instructions, preventing hidden optimization toward engagement and ensuring that their social media use aligns with their intended goals.

\vspace{-1mm}
\subsubsection{{Ease of use.}}
\textit{Users set and adjust feed goals without specialized vocabulary or interfaces.}
The interface between users and the feed algorithm should be simple, flexible, and expressive.  
Lowering the cost of articulation broadens \textit{who} can personalize feeds and reduces drift toward opaque defaults.  
This design goal draws on HCI systems (\S\ref{rw:control}) that show lightweight, accessible interventions are effective in helping people regulate their technology use~\cite{hinikerMyTimeDesigningEvaluating2016,gruningDirectingSmartphoneUse2023,koNUGUGroupbasedIntervention2015}.  
Ease of use is also important to ensure intent fidelity, as increasing the flexibility and ease with which users specify their intentions ensures alignment between intention and feed.

\vspace{-1mm}
\subsubsection{{Procedural transparency.}}
\textit{The feed creation process is procedurally transparent.} 
Users should clearly understand the process by which their feed is generated, and how different aspects of the configuration impact the resulting feed.
This design goal builds on research  (\S\ref{rw:transparency}) showing that people frequently misunderstand how feeds operate and consequently miscalibrate their trust or behavior~\cite{eslami2016first,rader2015understanding}, as well as on scholarship critiquing “black-box” notions of transparency and urging attention to the processes through which systems operate~\cite{ananny2018seeing}.  
Procedural transparency is essential to making the system easy to use: only when users understand how their input shapes the system's outputs can they refine and trust it.

\begin{figure*}[hbtp]
\centering
\includegraphics[width=0.8\linewidth]{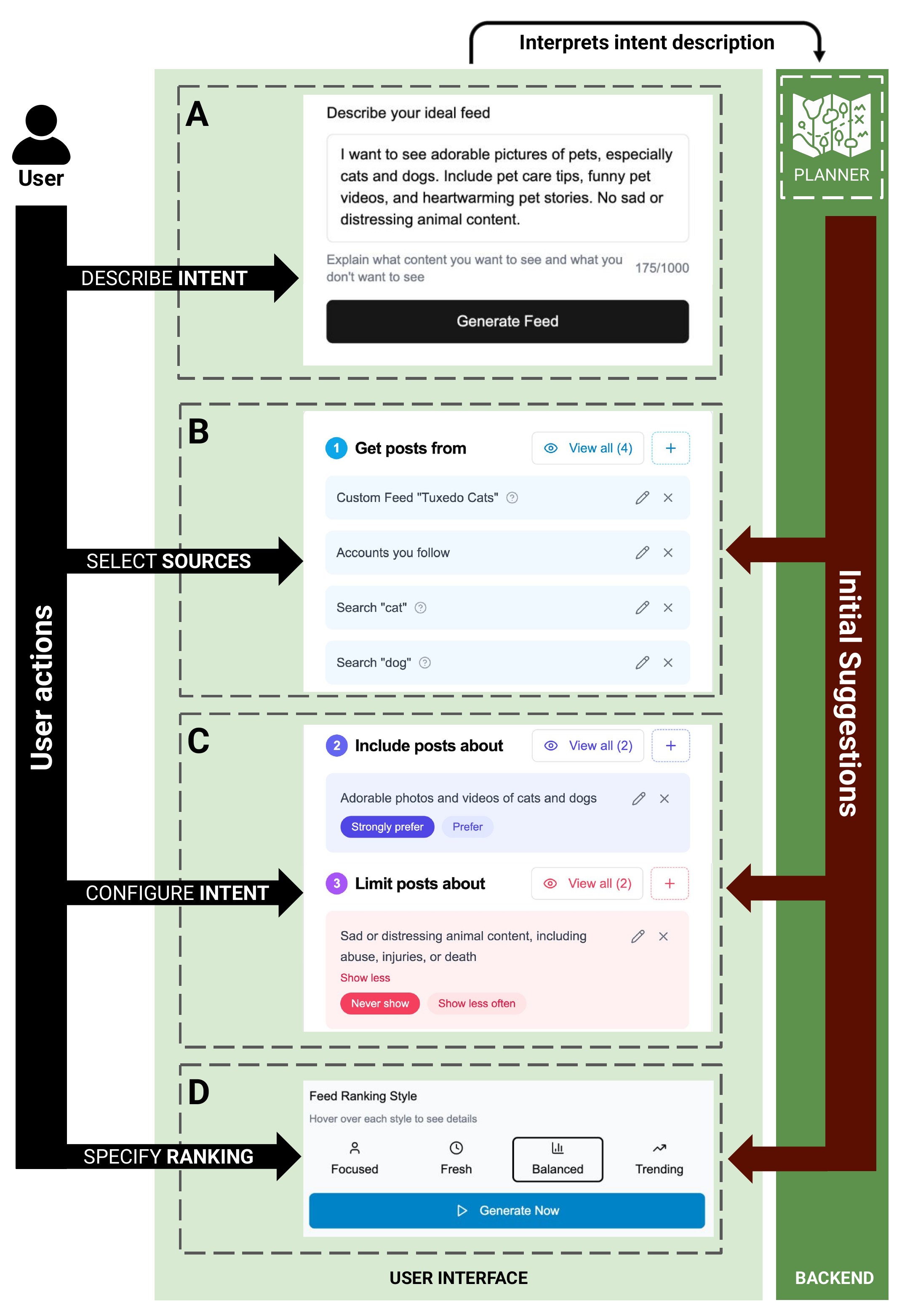}
\caption{Key components of the \System{} interface:  
(A) natural language input to describe feed purpose,
(B) source selection interface with suggested feeds, lists, and accounts,  
(C) preference management panel for including or limiting topics with importance levels, and  
(D) ranking presets balancing relevance, recency, and popularity.
}
\label{fig:interface-components}

\Description{This figure illustrates the interaction flow between the user interface and the backend in Bonsai, showing how users build intentional feeds step by step. On the left, user actions are broken down into four stages. In stage A, Describe Intent, users type a natural language description of their ideal feed, for example, asking for adorable pet pictures while excluding sad or distressing content. In stage B, Select Sources, the system suggests relevant feeds, accounts, and search terms, such as ``tuxedo cats'' or ``dog,'' which the user can edit or expand. In stage C, Configure Intent, users refine preferences into inclusion and exclusion rules: for instance, always including posts about cats and dogs, while limiting sad or distressing content. In stage D, Specify Ranking, users choose how their feed should prioritize relevance, recency, and popularity, with options like Focused, Fresh, Balanced, or Trending. On the right, the backend ``Planner'' interprets the user’s intent and generates initial suggestions that populate each stage. The process demonstrates how the system turns natural language goals into a structured feed configuration, balancing automation with user control.}
\end{figure*}

\subsection{Interface Design and System Walkthrough}

We designed \System{} based on the aforementioned landscape analysis and design goals.
During prototyping, we iteratively refined defaults and interaction flows in response to formative feedback from project researchers and informal demos with colleagues.
Two key changes from this process were: 1) the creation of an initial prompt interface (see \S\ref{sss:ipi}) to bootstrap the feed creation process; and 2) the addition of different ranking algorithms that prioritize recency, popularity, or topical fit (see \S\ref{sss:fc}).

We implemented \System{} for Bluesky for normative and pragmatic reasons. 
Bluesky is open, decentralized, built on the AT protocol, and designed to give users more control over their experience, content, and data. 
The AT protocol makes it easier to implement a feed curation system, and Bluesky users are also more likely to be interested in such an application.
The user experience (Figure~\ref{fig:interface-components}) unfolds through four steps, which we describe below.

\subsubsection{Describing intent}
\label{sss:ipi}
The workflow begins with a screen inviting users to describe their desired feed (Fig.~\ref{fig:interface-components}A).  
The text area includes a placeholder prompt and three example descriptions for inspiration.  
After entering their intent in natural language and clicking ``Generate Feed,'' the system parses the input and synthesizes it into a structured customization interface (Fig.~\ref{fig:interface-components}B--D).  

\subsubsection{Selecting sources}
Next, users specify the sources that will populate their feed (Fig.~\ref{fig:interface-components}B).  
Based on the initial description, \System{} suggests feeds, starter packs, lists, search queries, accounts, and hashtags.  
Users can accept, edit, or remove these suggestions, and manually add new sources.  
Sources can be accounts, other feeds, search terms, and `starter packs' (curated sets of accounts).

\subsubsection{Configuring intent}
In the third step, users refine what types of content should be prioritized or limited (Fig.~\ref{fig:interface-components}C).  
\System{} generates suggested prompts derived from the original description, which users can edit, delete, or expand with new plain-text entries.  
Each prompt is associated with a priority: under ``Include posts about,'' users can indicate whether posts should be ``strongly preferred'' or ``preferred,'' while under ``Limit posts about,'' they can specify that posts should ``never be shown'' or ``shown less often.''

\subsubsection{Specifying the ranking algorithm}
\label{sss:fc}
Finally, users control how posts will be ordered in the feed (Fig.~\ref{fig:interface-components}D).  
\System{} lets the user control the feed ranking algorithm by assigning importance to three metrics: relevance, recency, and popularity. 
The interface offers four presets: `Focused,' `Fresh,' `Balanced,' and `Trending,' each offering different trade-offs between the metrics. 
Hovering over the preset lets users see details about the importance given to each metric in the ranking algorithm (for more information on the presets, see the Appendix Fig.~\ref{fig:ranking-presets}).

\subsubsection{Generating and using the feed} 
After completing these steps, the user can click ``Generate Now'' to initiate the feed generation process. 
A few seconds after initiating generation, \System{} may suggest additional sources to the user, based on the edits made to the content preferences. 
After generation, the feed appears as a custom feed option in the user's Bluesky app.
\System{} continues to update the feed in the background.
If users wish to edit their feed, they can always make changes on the feed customization page.
They can modify preferences, add new exclusions, adjust source selections, and even create new feeds.

\subsection{System Architecture}

\begin{figure*}[ht]
    \centering
    \includegraphics[width=\linewidth]{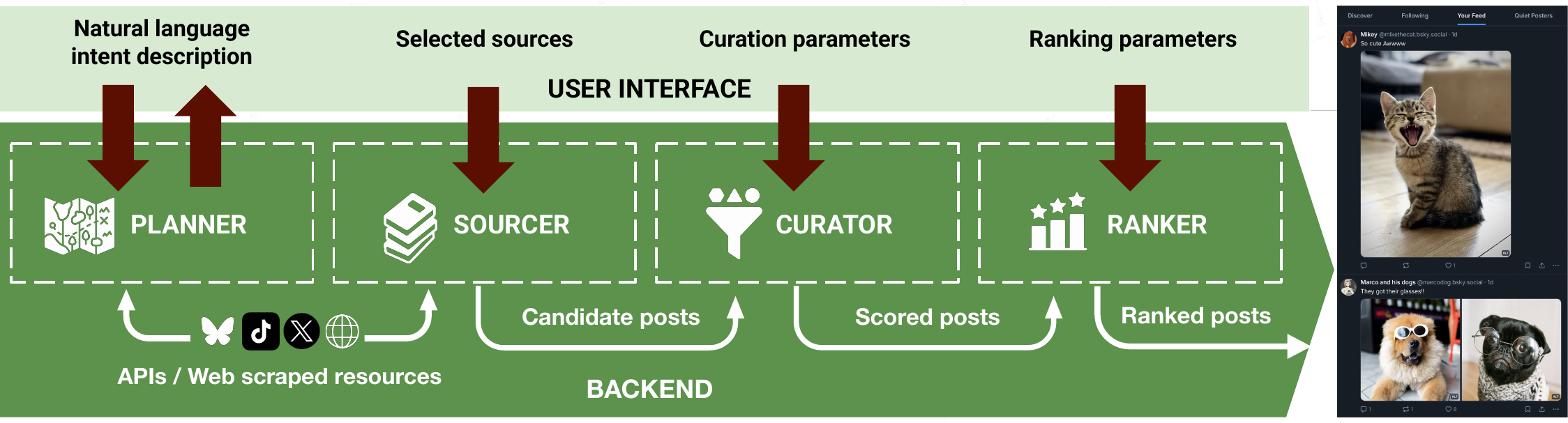}
    \caption{\System{}'s generalizable architecture for intentional feeds, composed of four components: Planner (translates natural language intent into an initial configuration), Sourcer (retrieves candidate posts from feeds, accounts, or hashtags), Curator (scores posts against user-stated preferences), and Ranker (orders content by relevance, recency, and popularity, as determined by the user). }

    \Description{This figure illustrates the system architecture of Bonsai, broken down into four main components: Planner, Sourcer, Curator, and Ranker. Together, they generate intentional and personalized feeds. At the top, the user interface begins with a natural language intent description, where users specify what kind of feed they want. This intent is processed by the Planner, which interprets the description and outputs an initial feed configuration. The Sourcer then identifies and retrieves content from relevant sources, such as accounts, hashtags, feeds, or search queries, utilizing APIs or web-scraped resources. These become the pool of candidate posts. Next, the Curator applies the user’s curation parameters, filtering and scoring posts according to inclusion and exclusion rules (e.g., ``always show pet videos,'' ``never show ads''). Finally, the Ranker orders the curated posts according to user-specified criteria such as relevance, recency, or popularity. On the right, the outcome is shown as a sample feed containing pet-related posts, demonstrating how the pipeline transforms user intentions into a live social media feed.}
\label{fig:system-overview}    
\end{figure*}

We propose a generalizable system architecture for generating \textit{intentional} and \textit{personalized} social media feeds. 
This architecture comprises four components (Planning, Sourcing, Curation, and Ranking), depicted in Fig.~\ref{fig:system-overview}. 
While our implementation targets Bluesky, the modular design is platform-agnostic: each component can be adapted to different ecosystems by plugging in platform-specific APIs, content databases, or models.  
We chose Bluesky because its open, decentralized AT Protocol makes external feed customization feasible.

\vspace{-1mm}
\subsubsection{Planning}
The Planning component translates a user's natural language prompt into an initial feed configuration.  
This includes suggested sources (e.g., feeds, lists, hashtags) and content preferences (e.g., ``include posts about'' and ``limit posts about'').  
Because this step involves semi-autonomous interpretation of natural language, we employ a language model (LLM) agent.  
In our Bluesky prototype, the agent queries an internal SQLite database of candidate sources (with previously scraped feeds, lists, and `starter packs') and generates search queries and filter suggestions from the prompt.
The resulting configuration is surfaced to the user for review and refinement.
Our implementation used GPT-4o via Azure; the Planning stage can be instantiated with other models or retrieval systems, depending on the target platform.  
We provide full details of the feed, starter-pack, and list database construction in Appendix~\ref{app:db-construction}, including our scraping procedure, filtering criteria, and the final counts (1{,}277 feeds, 205 starter packs, and 300 lists).
We provide additional details about the planner agent in Appendix~\ref{app:agent}.

\subsubsection{Sourcing} 

The Sourcing component retrieves candidate posts from the sources identified during Planning.  
In general, this stage serves as the bridge between user-defined inputs and the stream of available content on a platform.  
In our implementation, we used the Bluesky AT Protocol APIs to collect posts from feeds, accounts, and queries.  
For the initial generation of the feed, we extract all posts from the past 96 hours from each of the sources, up to 100 posts per source, since the goal of the initial generation of the feed is to provide a representative starting point for the user to consume.
For ongoing operation, the system refreshes feeds every 30 minutes by fetching new posts since the last update.  
This logic can be adapted to any platform that supports source-level queries or APIs for bulk content retrieval.  

\vspace{-1mm}
\subsubsection{Curating}
The Curating component applies the user’s stated preferences to the sourced posts and assigns each item a relevance score.  
In principle, this stage can be implemented with classifiers or rules tailored to the platform; in our case, we leveraged a multimodal LM (GPT-4o) to handle Bluesky’s mix of text and images and to align directly with users' natural language prompts.  
Each post received a score on a 0--10 scale: 
``strongly prefer'' (8--10), 
``prefer'' (5--7), 
 ``unspecified'' (3--4).  
``show less often'' (1--2), ``never show'' (0).
Posts scored as zero were filtered out.  
Although our prototype relied on a relatively heavyweight model, the curation logic could be reimplemented with lighter models, heuristics, or hybrid pipelines, depending on deployment constraints. We provide the full prompts used in Appendix~\ref{app:prompts}.

\vspace{-1mm}
\subsubsection{Ranking}
The Ranking component orders the curated posts into a final feed according to user-specified priorities.
We adopted three widely used dimensions from existing feed algorithms: relevance, recency, and popularity~\cite{covington2016deep,guessHowSocialMedia2023}.  
Relevance scores came from the Curating stage; popularity was computed as a weighted engagement score (likes + 3 $\times$ reposts + 2 $\times$ replies); and recency was calculated via reverse-chronological order.  
To combine these, we applied a Weighted Borda Count, with weights set by the user’s chosen ranking style (e.g., ``Focused,'' ``Fresh,'' ``Balanced,'' ``Trending'').  
This structure generalizes readily: other platforms may substitute alternative engagement metrics, quality signals, or normative objectives.  
We provide more information and the implementation of the Weighted Borda Count algorithm in Appendix~\ref{app:borda}.

\vspace{-1mm}
\subsubsection{Backend}
We implemented \System{} as a web application with a Next.js/React frontend and a Python/Flask backend, hosted on Azure.  
The version of \System{} presented in this paper only supports Bluesky's AT protocol but is designed to be extensible to other platforms as long as platform-specific APIs and/or Web scraped resources are available or constructed. To log in to our system, users can log in using their Bluesky account with an `app password'. The user feeds and configurations are stored in a SQLite database on the same machine as the Python backend. When a user makes a request through the Bluesky client, \System{}'s backend returns their current, active feed. 
We used OpenAI GPT-4o hosted on Azure as the LM provider, motivated by credit availability at our institution. 
Yet, the architecture is designed to be model-agnostic: different LMs or even self-hosted models can be substituted, and the system can be extended to additional platforms beyond Bluesky.

\begin{table*}[ht]
\centering
\small
\renewcommand{\arraystretch}{1.35}
\caption{Participant demographics, Social media usage, and \System{} usage statistics. `Custom' in `Bluesky feeds used' can represent any custom feed existing on Bluesky, made by any user. `Manual gen.' stands for the manual regenerations of a user's feeds. `Weekly feed loads' represent the number of times users loaded the feed through their clients per week, on average.}
\label{tab:participants}
\begin{tabular}{
@{} l l l >{\raggedright\arraybackslash}p{3.8cm} >{\raggedright\arraybackslash}p{1.6cm} |
S[table-format=1.0]
S[table-format=2.0]
S[table-format=3.0]
@{}}
\toprule
\multicolumn{5}{c}{\textbf{Demographics \& Social Media Usage}} & \multicolumn{3}{c}{\textbf{\System{} Usage}} \\
\midrule
\textbf{ID} & \textbf{Country} & \textbf{Bluesky use} &
{\makecell{\textbf{Social media}\\\textbf{platforms used}}} &
{\makecell{\textbf{Bluesky}\\\textbf{feeds used}}} &
{\makecell{\textbf{Feeds}\\\textbf{built}}} &
{\makecell{\textbf{Manual}\\\textbf{gen.}}} &
{\makecell{\textbf{Weekly}\\\textbf{feed loads}}} \\
\midrule
P1  & US      & Once a week           & Twitter/X, Bluesky, Instagram, Youtube, Facebook                            & Following & 2 &  5 &  36 \\
P2  & US      & Once a week           & Twitter/X, Bluesky, Instagram, Youtube                                      & Following & 2 &  7 & 104 \\
P3  & US      & Once a day            & Twitter/X, Bluesky, Instagram, Youtube, LinkedIn                            & Following & 3 &  7 &  41 \\
P4  & UK      & Several times a day   & Twitter/X, Bluesky, Instagram, Youtube, Facebook, LinkedIn                  & Following & 3 &  5 &  68 \\
P5  & US      & Several times a day   & Twitter/X, Bluesky, Mastodon, Instagram, Youtube, Facebook, Tiktok          & Following, Discover, Custom & 3 &  5 & 249 \\
P6  & US      & Once a day            & Twitter/X, Bluesky, Mastodon, Instagram, Threads, Youtube, Facebook, Tiktok & Following, Discover, Custom & 2 &  2 &  53 \\
P7  & US      & Once a day            & Twitter/X, Bluesky, Instagram, Threads, Youtube, Tiktok, RedNote            & Following, Discover, Custom & 2 &  6 &  42 \\
P8  & Austria & Several times a day   & Bluesky, Instagram, LinkedIn                                                & Following & 1 &  9 &  90 \\
P9  & UK      & Several times a week  & Twitter/X, Bluesky, Youtube, Tiktok, Linkedin, RedNote                      & Discover & 2 &  8 & 126 \\
P10 & US      & Several times a day   & Bluesky, Youtube, Substack                                                  & Following & 2 & 13 &  56 \\
P11 & US      & Several times a day   & Bluesky, Youtube, Twitch, Reddit                                            & Following, Discover, Custom & 4 &  5 & 264 \\
P12 & US      & Once a day            & Twitter/X, Bluesky, Instagram, Tiktok                                       & Following & 2 &  8 &   8 \\
P13 & US      & Several times a day   & Twitter/X, Bluesky, Instagram, Youtube, Tiktok                              & Following & 1 &  2 & 178 \\
P14 & Canada  & Once a day            & Twitter/X, Bluesky, Instagram, Facebook                                     & Following & 4 &  5 &  69 \\
P15 & Spain   & Several times a day   & Bluesky, Discord                                                            & Following, Discover, Custom & 3 & 16 &  31 \\
\bottomrule
\end{tabular}

\Description{
This table summarizes participant demographics, social media usage patterns, and their use of Bonsai. The left side shows Demographics \& Social Media Usage. Each row corresponds to a participant (P1–P15), with columns for their country, frequency of Bluesky use (ranging from once a week to several times a day), and the other social media platforms they regularly use (e.g., Twitter/X, Instagram, YouTube, TikTok, Mastodon, Reddit, Discord, LinkedIn, Substack). The right side shows Bonsai Usage. Columns indicate which default Bluesky feeds participants used (e.g., `Following,' `Discover,' or `Custom'), the number of feeds they built during the study (from 1 to 4), the number of times they manually regenerated their feeds, and their average number of weekly feed loads (from as few as 8 to as many as 264). Overall, the table highlights that participants varied widely: some used Bluesky only once a week and built just a couple of feeds, while others were heavy social media users who experimented with multiple Bonsai feeds, frequent manual regenerations, and very high weekly feed loads. This variation underscores the diversity of engagement styles in the study.
}
\end{table*}

\section{Methods}

We conducted a field study with 15 participants who used \System{} for a median of 11 days (\textit{min}: 6 days; \textit{max}: 25 days; \textit{avg}: 12.27 days). We aimed for approximately one week of use for each participant, though the exact number of days varied based on availability. We
interviewed participants at the beginning and end of the study in order to understand the details of their experience. The study lasted one month, from July 29th to August 29th, 2025.

\subsection{Participants}

We recruited participants primarily through social media, creating a visual flyer that we shared on Bluesky and in relevant Slack groups.
We also conducted targeted outreach on Bluesky to engage frequent users of the platform and to enlist non-academics in the study.  
From those who filled out the interest survey, we contacted participants we could validate as real people and as regular Bluesky users.  
In total, 15 participants enrolled in the study.  
Participants received a \$50 gift card for completing both interviews, and all procedures were approved by Princeton IRB (\#18334).
Throughout this paper, we anonymize participant names using \textit{PX} nomenclature and further obscure highly specific stories or feed intentions that could be recognizable.

We took a broad stance on participant inclusion: the only requirement was that participants be 18 or older and actively use Bluesky. Despite this broad criterion, likely due to the skew of our own outreach, all participants except one were academics. 
We purposefully sought participants with different levels of Bluesky and custom feed usage, and five participants already used other custom feeds on Bluesky.
Eleven participants lived in the US, two in the UK, and the others in Canada, Austria, and Spain.

\subsection{Procedure}

The study comprised an onboarding interview, at least a week of using \System{} feeds, and a follow-up interview.  Both interviews were semi-structured and lasted 45 minutes. 
In the first interview, we explored participants’ general social media habits, asking what value they currently derive from feeds, what they feel is missing, and what they wish feeds would provide.  
We then walked them through the feed generation process in \System{}, asking them to think aloud as they selected sources, wrote prompts, and configured ranking options.  
During the trial week, participants were encouraged to use \System{} in their everyday browsing, creating and refining feeds as they wished.  

The second interview took a more reflective and evaluative form. 
We asked participants to describe how they attempted to convey their intent in the feed, whether they felt successful, and what challenges they encountered.  
We also probed the broader pros and cons of intentionality, natural language input, and feed-level control, as well as participants’ overall impressions of the system.  
In several cases, we revisited the feed creation process to observe how their strategies evolved.  
In both interviews, we examined the feeds participants had generated, discussing whether the content matched their expectations and probing moments when posts seemed out of place. 
We include the full interview protocols in the Appendix sections \ref{interview-protocol-1} and \ref{interview-protocol-2}.

\subsection{Data Analysis}
Participant interviews were recorded and automatically transcribed on Zoom.
We uploaded the anonymized transcripts to Atlas.ti, where they were further refined for accuracy.
The first and second authors then engaged in open coding for two full interviews each, discussing and honing the initial codes with the broader research team.
As additional interviews were being conducted, we iteratively expanded the codebook and discussed emerging themes. Some of the most relevant initial higher-level themes we identified included \textit{social media motivations and goals}, \textit{relationship between control and agency}, \textit{describing intent in natural language}, and \textit{feed curation effort}. Each of these had several subcodes, for example, \textit{describing intent in natural language} encompassed \textit{preferences hard to capture in natural language}, \textit{perceptions and folk theories of LLMs}, \textit{intuitive}, \textit{powerful}, and \textit{other}. This codebook and discussions with the research team guided interviews, for example, focusing more on effort and transparency since these were yielding compelling insights from participants.
We continued to iterate on these themes through discussions as interviews were conducted, eventually uniting them into three main overarching sections which align with our design goals: \textit{imprinting intentions on social media}, \textit{controlling intentional and personalized social media feeds}, \textit{procedural transparency of intentional and personalized feeds.}

In all, we conducted fifteen interviews. Interviews continued until thematic saturation was reached (no new codes added after two interviews) after fourteen interviews. We considered this sample size sufficient given prior studies identified that the mode number of participants in HCI studies is twelve~\citep{caine2016local}. Nonetheless, we conducted one additional interview to capture more non-academic perspectives. 
We note that, given our small and predominantly academic sample of active Bluesky users, our findings are not intended to be statistically generalizable, but to characterize how this particular population appropriates and experiences LM-mediated intentional feeds.

To add richness to our findings, we also captured the \System{} interaction logs and the natural language prompts from participants.
The natural language prompts were tagged with specific user requests, such as a topical focus or a request to filter specific content.
Additionally, we manually analyzed the logs for each participant to identify the actions they took on the platform between interviews, such as editing a feed or creating a new feed.
These results are incorporated throughout the findings section.

To ensure that \System{}-generated feeds were adequate, we had 7 annotators each annotate 60 posts. For each post, annotators were given user-defined prompts and asked to judge whether posts were "perfectly relevant," "relevant," "not relevant," "should have been excluded," or "not sure." Each annotator annotated posts for two different feeds (30 each).
70\% of posts were rated as relevant or better (26\% "perfectly relevant," 44\% "relevant"). Further,  16\% were rated "not relevant," and 7\% "should have been excluded." We consider these annotations evidence that the system can generate feeds that largely align with users' stated intentions, especially given that the 14 considered feeds were randomly selected (and not necessarily the best-performing ones).
We had an eighth annotator label posts that had been previously annotated, finding substantial inter-annotator agreement ($\kappa = 0.77$).

\section{Results}

We discuss how participants used \System{} during the study period, focusing on the prompts they authored, the actions they performed, and the ways they evaluated their feeds. We structure our findings around our three design goals: (1) how participants \textit{imprinted their intentions on social media feeds}, (2) how they \textit{steered and maintained intentional and personalized feeds}, and (3) how they made sense of the \textit{procedural transparency of feed generation}. 
Our 15 participants built between 1 and 4 feeds each for a total of 36 feeds. They manually re-generated their feeds between 2 and 16 times for a total of 103 times. Per week, they consumed \System{} feeds between 8 and 264 times each for a total of 1413 requests. 

\subsection{Imprinting intentions on social media feeds}

We describe how participants used ~\System{} to create and reflect on intentional and personalized social media feeds.
Largely, participants used \System{} to define broad and specific topical interests, filter out irrelevant and harmful content, and to flexibly convey various other structural intentions.
We also include participant reflections on emergent vs intentional feeds, and how using natural language felt for conveying intentions.

\subsubsection{Participants specified topical interests using \System{} to create both general and specialized feeds.}
Some participants tried to construct a ``default'' feed that would encompass all their primary interests.
For example, in 20 out of the 36 feeds constructed by participants, two or more distinct topics were included. 
P8's instructions mentioned they wanted to see \textit{``new research on quantum computing''}, \textit{``developments in Machine Learning research,''} and \textit{``politics with a special focus on Europe''}.
P10's instructions aimed to improve their current feed by specifying interests \textit{``not represented by [their] current feed like: music, biking, meditation, architecture.''} 
For many, this urge to discover new content seemed to stem from issues with the Bluesky platform itself.
P3 said \textit{``the engagement-based content, the }`For you'\textit{, I feel like it just doesn't work [...] because it shows me a bunch of stuff that I really don't care about.''}
These participants viewed \System{} as a way to produce `better' feeds than their main, central feed on Bluesky or get exposed to content they otherwise would not be.
In their second interview, P6 said their \System{} feed \textit{``was surfacing stuff that I wasn't finding in other places''}, while P8 was happy that they
\textit{``got a lot more diverse stuff [and] got articles from newspapers I usually don't follow.''}
P11 said that their feed was \textit{``bringing me posts from people, accounts, and places that otherwise I never would have encountered, and that's the intent [...] to see things that I wouldn't otherwise encounter.''}
Most enthusiastically, after generating their first feed, P13 said
\begin{quote}
\textit{``This is exactly what I'm talking about. I didn't know this was on here. This is the kind of thing that I want to see. (...) This is the most interesting my feed has been in a while.''} -P13
\end{quote}

In the remaining 16 feeds, participants used \System{} to build feeds for specific topics or hobbies they were interested in. 
P7 created a feed focused on cinematography (\textit{``I want to see film festival news, film critics, artists' sharing their experiences or new releases, experimental / AI films''}), P11 on college sports (\textit{``I want to see all posts related to US college basketball''}), and P15  (\textit{``a videogame development feed.''})
Before using \System{}, some participants had already used complementary topical feeds, such as P6, who followed a feed of political science journals, while P9 would repeatedly search for common keywords by clicking on the same keywords from their saved search history -- effectively creating a pseudo-feed.
For this participant, \System{} offered a much easier way to create a topical feed suited to their interests.

Regardless of whether participants wanted a feed with one or multiple topics, they often leveraged the flexible LLM-driven nature of \System{} to specify complex sub-topics or go beyond surface-level feed topics.
In this vein, P15 created a feed that was specifically about video game development and building, rather than a more general video game feed; while P7 wanted to see news about experimental/AI films as opposed to commercial movies.
For participants who were already relying on topic-based feeds, \System{} offered a way to hone in on specific sub-parts of a topic. For example, P11 followed sports feeds already but wanted some college basketball teams to be featured more heavily than others. These kinds of within-topic customizations had been hard for participants to achieve before \System{}: \textit{``this [feed] was really, really good, this is exactly what I wanted [\ldots] it was a lot easier to dial this one in'' (P11).}

\subsubsection{Participants used \System{} to filter out irrelevant and toxic content and to improve productivity and well-being.}

On 15 out of the 36 feeds, participants included specific instructions to filter out or reduce some type of content.
The first instance when participants wanted to filter out content was when they found it irrelevant. 
For example, P4 wanted to filter out US-centric topics he was not interested in: \textit{``I want my chronological following timeline without any US mention.''}
The second use case for filtering out content on \System{} was to try to limit negative posts that could affect participants' personal well-being.
To that end, P6 wanted their feed to \textit{``avoid outrage-farming content, repetitive posts, and really political takes''}, and P15 wanted to \textit{``avoid posts containing news, sad, rude or controversial content, quote-dunking, politics, discussions, or polarizing content''}.

These desires for better filters stemmed from participants' negative social media experience, particularly on Twitter. 
Some users sought to improve their mental health on platforms like X, which P8 called \textit{``toxic''} and \textit{``unbearable,''} P10 reported feeling \textit{``anxious when using it,''} and P6 felt made them \textit{``deeply unhappy.''}
Others felt that engagement-based feeds simply were not effective at filtering out content, even when given explicit instructions.
P8 said that on Twitter they \textit{``would always say, okay, I want to see less of that, and I wouldn't see less of it,''} while P15 held the same opinion of Bluesky, where \textit{``even though you tell it `show less of this', it keeps displaying it.'' }
P4 told the story of a user whose academic content they enjoyed, but who also posted frequently about football---a topic they had no interest in.
In response, P4 tried to block football-relevant keywords, but found  \textit{``with the existing methods, you can achieve certain results, they are certainly useful in some contexts, but they're not very specific, they're not very, sort of fine-grained.''} 
P13 shared a similar experience, saying \textit{`on Bluesky, I would say [filtering] is terrible. I have set up 30 keywords to mute things, so that I don't see political content on Bluesky, but it still manages to creep onto my feed.'' }

In that context, participants were satisfied with their experiences filtering undesired content on \System{}. P13 said that
\begin{quote}
\textit{``A lot of people I follow also like to post politically [...] but I want to see their research posts, I don't want to see their political posts, and this feed [...] allowed me to see the things I liked from the creators, and not the things that I was not interested in.''} -P13
\end{quote}
Similarly, P14 said that their favorite thing about \System{} was that it `eliminated junk' from their feed, and that filtering `was working perfectly' for their feeds.
For these users, the flexibility of \System{} allowed them to avoid exposure to uninteresting posts or interacting with potentially harmful content.

\subsubsection{Participants flexibly used \System{} to specify a variety of other structural intentions.}
Participants often added structural intentions to their instructions.
Of the 36 created feeds, 23 included a preference for specific content and media formats, such as P13, who requested \textit{``arxiv links''} and \textit{``github repos for code.''}
Sixteen feeds included criteria for the sources that should appear in the feeds, such as P11, who asked for \textit{``individual scholars, department accounts, professional associations, and all other relevant institutional accounts''} and to \textit{``minimize posts from accounts with fewer than 500 followers.''} 
Fourteen feeds included language and geographic restrictions, like P15 who wanted a \textit{``Spanish-speaking only feed''} or P10 who wanted content from \textit{``people in my local area.''} 
Twelve feeds included quality and engagement criteria (P3: \textit{11 sources that are trustworthy, respectable, and neutral (a bit of liberal bias is ok, but I don't want conservative outlets)''}), ten feeds specified temporal restrictions (P14: \textit{``recent political happenings and news''}; P15: \textit{``in the last hours''}), and six feeds included relational aspects with the user's existing network (P10: ``new accounts that are followed by people I currently follow.'')

\subsubsection{Participants felt liberated using intentional feeds, though some already felt they could convey intentionality through behavior in engagement-based feeds.}
Participants were intrigued by the prospect of crafting an intentional social media feed. Participants noted how foreign this concept felt, like P10, who explained \textit{``normally, when I use Twitter or Blue Sky or something, I kind of have some implicit ideas, but I never try to say it explicitly, [...] so just trying to be intentional that way is already interesting.''}
Yet, participants found that explicitly defining feed intent was liberating, compared with engagement-based paradigms.
P6 mentioned that using more intentional instructions helped them detach their intentions from their engagement behavior:
\begin{quote}
\textit{``I don't have to be selective about my behavior [when using \System{}], where a like, a reply, a repost, to me, is communicating something different than }`I want to see this more.'\textit{ [...] it's very nice for me to be able to perform those actions without having to be conscious about }`oh, this may be detrimental to my future experience on this platform.'\textit{''} -P6
\end{quote}

Similarly, P8 said that engagement-based feeds can be misleading, as \textit{``sometimes you just stay on a post longer time, and then, just like your full feed is based on that, even though maybe you weren't even that interested in it.''}
P11 found engagement-based feeds draining and manipulative, referring to \textit{``that algorithmic rabbit hole [...] constantly trying to feed and gauge your attention.''}
For these users, a feed that was purely based on intention---not engagement---helped to ease the cognitive burden of knowing that their behavior could impact the content they would be served.

A second way in which \System{}'s intention-based feeds felt liberating to users is that they eliminated the pressure to follow certain accounts.
Participants stated that the choice to follow a user can be socially motivated, rather than being driven by a desire to view a user's posts.
P1, for example, explains that for \textit{``academic follows [...] If I don't follow certain people, I'm being a bad member of the community, or something like that.''}
P3 explains that there is a tradeoff between social expectations of following an account and the impact a follow can have on their feed, saying they  \textit{``don't use follow as a 'showing support tool', which would be the case if it didn't affect necessarily what I see.''}
Thus, \System{} also helped to relieve the social pressures of following specific accounts.

Although participants felt intentional feeds could be liberating, some participants felt perfectly in control of their existing engagement-based feeds.
P13 for example explained \textit{``on Twitter, [...] I could just hit 'not interested,' and then I won't see stuff,''} and that they feel \textit{``I have a lot of self-control on what I click on to shape my algorithmic feed, and so because of that, I can get a really good feed on Twitter.''}
P9 similarly felt confident that platforms could infer their intent correctly: \textit{``once I give somebody a thumb [or a] heart, [the platform] can more quickly capture my interests and then start recommending more things similar to that''}.
Even those who felt more lukewarm about their ability to manage engagement-based feeds still exerted additional controls over their feeds by blocking and muting accounts and keywords to filter out content they did not wish to see.

\subsubsection{Natural language is a promising way to define feed intentions and preferences, but not all feed intentions lend themselves to natural language.}

Through our design, participants cleverly used natural language to convey specific and nuanced desires.
For example, participants used humor and specific cultural references to help the AI ``situate'' itself within the type of desired content.
Nuanced references could allow participants to convey complex semantic ideas with just a few words.
These specifications served as shortcuts for ``vibe preferences'' (based on ``vibe coding'' ~\cite{roose_2025_not}); where users did not fully specify what they were looking for, but challenged the system to interpret rich input.
Participants largely felt starting with an LM prompt felt like \textit{``a very natural way to get [the intention] down at first''} (P6).
P4 also conveys this elegance:
\begin{quote}
\textit{``It is more elegant, because you can sort of craft in your own words what you desire, and you don't have to rely on various crude, and coarse measures such as blocking, or muting, or sort of eliminating posts based on keywords, that you don't want to sort of see. Like, that all works to different levels, and depending on what you want, might even work perfectly.''} -P4
\end{quote}



On the other hand, participants also pointed out specific times they could struggle or even feel uncomfortable voicing their preferences in natural language.
P10 noted this tension from the get-go: \System{} \textit{``made me realize that it can be difficult to articulate in natural language our specific interests.''}
P8 struggled with defining a certain type of post they did not want to see: \textit{``because it was not an advertisement, it was just, like, someone saying, hey, this is a tool''} and lacked \textit{``the general absence of a clear definition on what it was that I didn't want to see.''}
When making edits, P8 stated \textit{``I think it was because I was not really sure about what I was even seeing, and then having to communicate that [felt challenging].''}
P6 specifically tried to describe a feed \textit{``personalized to my specific humor''} and found that \textit{``pretty hard to do.''}
One participant said they would prefer to see less of a specific political conflict that negatively affected their mental health, but felt that writing this desire out in words felt ``wrong,'' since others had to experience the conflict and could not tune it out.
P1 also felt it was ethically complex to intentionally define the types of news they want to see, saying that although they are \textit{``always creating an echo chamber [in implicit feeds], it just feels bad to explicitly write these rules.''}

Due to these constraints in expressing preferences in natural language, participants sought to combine \textit{intentional} and \textit{emergent} preferences.
For example, some participants mentioned that they would benefit from providing examples of posts they liked/disliked without having to describe them in natural language.

\begin{quote}
    \textit{``There's that famous Supreme Court case where the justice said: `I can't define porn, but I know it when I see it.'\footnote{From ``I know it When I See It:'' A History of the Definition of Obscenity: \textit{``I shall not today attempt further to define the kinds of material I understand to be embraced within that shorthand description ["hard-core pornography"], and perhaps I could never succeed in intelligibly doing so. But I know it when I see it, and the motion picture involved in this case is not that''}.} I think there's a general sense, where I think even when I'm trying to describe what my current feeds are like, and then what my preferred ones are like... [...] I find those very challenging to describe and put words to. I think it's more of an experience.''} -P5
\end{quote}

\subsection{Steering social media feeds}

A central focus of \System{} was to give participants a simple way to construct personalized feeds.
Our findings show that while participants valued this agency, they also grappled with the effort required to maintain it.  
This tension between effort and {reward} emerged as a recurring theme: some participants found the work of curating feeds worthwhile for the increased precision it afforded, while others felt the costs outweighed the benefits once their feeds reached a ``good enough'' state.  
Participants with prior feedbuilding experience found \System{} comparably easy to use, while those used to emergent feeds could struggle with the magnitude of the paradigm shift.

\subsubsection{Natural language reduces the startup effort to obtain relevant content.}
Overall, participants were pleased with the initial ease of creating their first feed. 
Participants saw how their descriptions immediately yielded relevant feed settings, for example, P8 felt \textit{``[the feed] is very accurate to what I described.''}
After they first looked at their feed, P6 claimed \textit{``I'd be happier with this than my base feed.''}
To P3, \textit{``the initial step of getting a feed that can interest me, and getting something that works, was very quick and empowering, so it was a nice experience to be able to do that.''}

Participants with experience with other feedbuilders specifically mentioned how much easier it is to use \System{} compared with other feed building tools available for Bluesky, like \textit{graze.social} or \textit{Skyfeed}.
P6 enjoyed that \System{} easily facilitated setting up a feed, and observed that setting up custom feeds through \textit{Skyfeed} or \textit{graze.social} seemed like \textit{``a lot up front to not even know if you're gonna get a particularly good feed out of it.''}
P15 has a lot of experience using feed building systems, and said \textit{``it would have taken me probably hours in between creating, backfilling it, putting it in production, and then fine-tuning it to work. In here, in five minutes, you already have something that is a super good, almost final product.''}
Specifically, the other feed building systems available to use are all rule-based.
P11, who had used Skyfeed in the past, explained that in other interfaces they were \textit{``just basically writing Boolean operators about these combinations of words,''} whereas \System{} let them \textit{``just kind of stream of consciousness talk about the things you want to see, which is much more intuitive, right? It gets you closer to that--- I know what I want to see, rather than trying to sit there and think of keywords.''}
Similarly, P15 described using \System{} as \textit{``super natural:''} \textit{``doesn't require you to go into logic mode and start thinking about the building process by itself.''}
In all, compared to current state-of-the-art feed builders, participants described \System{} as easier to use, particularly at setup.

\subsubsection{User-controlled social media feeds can require significantly more effort to maintain.}

After the initial feed creation process, not all participants had the same motivation to keep editing their feed.
Overall, the amount of control and enjoyment that participants felt when building and maintaining feeds with \System{} seemed correlated to the amount of effort they subsequently invested.
On one side, four participants chose to put in significant effort to try to improve their feeds and reported that this effort was worth the improvements.
P11, for example, took \textit{``an extra 5 minutes''} to manually add a subset of accounts to the list of sources after the initial suggestions, and describes their experience:
\begin{quote}
\textit{``It was completely worth it, because that sharpened the feed dramatically once I made sure those [accounts] were included. Being able to add these specific accounts and be that granular was incredibly important to me (...) being very hyper-specific like this would help me get as close to the feed as I want to as soon as possible.''} -P11
\end{quote}

Five participants were satisfied with their initial feed, made minimal edits, and then estimated that additional effort was not worth improvement in their feeds.
P6 is one such lightweight user. Once they found their feed useful enough and hadn't encountered any \textit{``deeply upsetting''} content, they did not feel the need to keep tweaking it, saying \textit{``I think the effort to reward or change ratio no longer made sense, [...] I wouldn't go in to be like, `okay, I don't want job posts'.''}
P8 enjoyed that with the natural language interface \textit{``if I can just put text, it's like, okay, I [wrote] two sentences, or let it be 10 sentences, and then I have, kind of my preferences captured.''} Yet, P8 felt that further fine-grained edits were unnecessary, saying \textit{``having too much individual control is too much effort, [...] I don't think I would want to put that much effort into engineering my feed.''}

Finally, for the six remaining participants, their initial feed did not meet their expectations, and they ultimately either stopped using it or created a new feed where they had more success. This was the case for P10, who prompted the system to suggest \textit{``new accounts that are followed by people [they] currently follow''} and asked for \textit{``more users local to [them]''}. 
However, the current system is not equipped with account suggestions based on existing accounts and lacks data on the location of accounts or posts.
While the system still tried to suggest relevant content to the user, future iterations should more clearly communicate system constraints.
Due to these restrictions, P10 built a final feed centered around a niche topic of print media and zines and found this feed \textit{``more enjoyable.''} 
They explained that their first feed \textit{``felt overwhelming a bit, because everything seems totally new from a person that I didn't know.''} 
Another example is P3, who tried recreating a feed similar to their `following' feed but noticed that \textit{``the content that I engage most with from this feed is coming from the accounts I already followed''}, which led them to prioritize continuing to use their 'following' feed instead of their \System{} feed.
P3 explained that they felt the effort trade-off was not worth it with the time they had: \textit{``it's a hard trade-off, because time is both what I gain, but also what I would need to put in to make it better.''}

\begin{figure*}[t]
    \centering
    \includegraphics[width=0.925\linewidth]{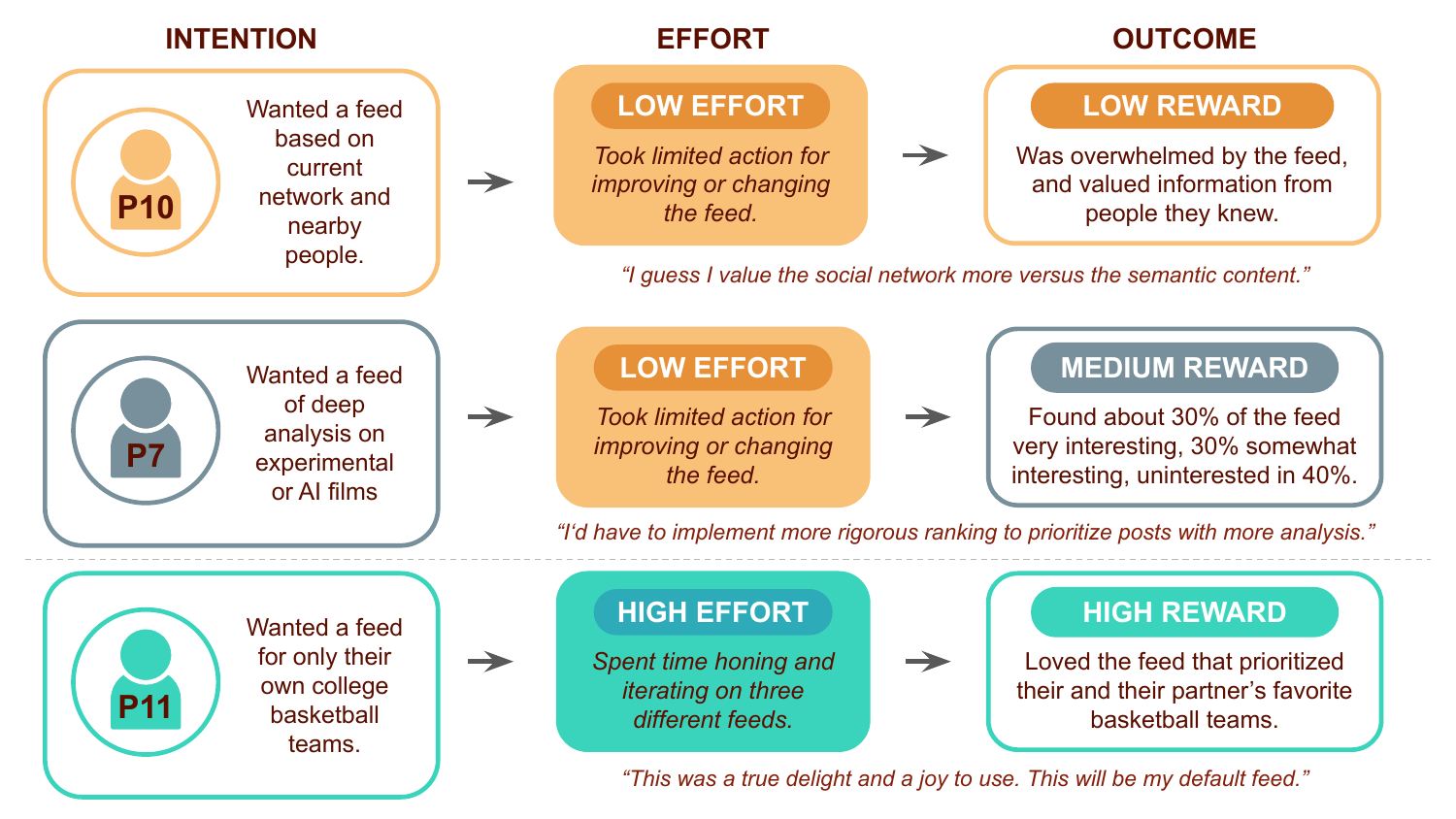}
    \caption{Three representative participant user journeys using \System{}: low effort, low reward (top row); low effort, medium reward (middle row), high effort, high reward (bottom row). The top two participants (P10 and P7) represent perspectives of people who spent minimal effort to improve their feeds. P11 (bottom) represents a participant who had prior experience using other feed builders, and invested time to optimize their \System{} feeds and reap high rewards.}
    \Description{A diagram showing three user scenarios for feed customization, each with Intention, Effort, and Outcome columns. Participant 10 (orange row) wanted a feed based on their current network and nearby people, exerted low effort, and experienced low reward, feeling overwhelmed and valuing social connections over content. Participant 7 (gray row) wanted deep analysis on experimental or AI films, exerted low effort, and achieved medium reward, finding 30\% very interesting, 30\% somewhat interesting, and 40\% uninteresting. Participant 11 (teal row) wanted a feed for only their college basketball teams, exerted high effort by iterating on three different feeds, and achieved high reward, enjoying the final result.}
\label{fig:user-journeys}    
\end{figure*}

Participants thus had three different types of experiences, and generally achieved more rewarding feeds when they put more effort into learning and updating the \System{} system. We illustrate these divergent types of user journeys in Figure~\ref{fig:user-journeys}. 
Though we cannot draw any definite quantitative conclusions, our interviews suggest that the two reasons that motivated participants' willingness to invest additional effort were prior experience with custom feed builders and a strong desire for specific feeds.
P13 had never used other feed builders, but was dissatisfied with Bluesky's default `Discover' feed, and was therefore highly motivated to create a new feed, finding that their \System{} feed is \textit{``way better than the Discover feed on Blue Sky.''}
Prior experiences with feed-building tools also separated participants who put in extra time to refine their feeds from participants who did not.
As we highlight in Section 5.2.1, those who were used to other feedbuilders were accustomed to effortful feed creation experiences, and appreciated the comparative flexibility and ease of using \System{}.
In contrast, participants with limited experience with other feed builders brought only their expectations from emergent feeds, which require minimal effort. 
For these participants, spending time constructing an intentional feed represented a paradigm shift that did not align with their expectations.
P10 captured how intentional feed editing can seem quite intimidating for novice users: \textit{``it feels like there's a bit of a learning curve for how to use this tool, because [\ldots] we're expecting to be fast when we use [feeds].''}

\subsubsection{Participants reported they felt empowered but struggled with increased cognitive load.}
Despite the effort required to create feeds that fully meet user intentions, some participants in our study felt empowered and excited by the ability to curate their feed.
P4 said \textit{``it is empowering to actually have a different tool in the toolbox, which you can use to create, sort of, your own... Custom universe, if you will.''}
Similarly, for P11 \textit{``this was a true delight and joy to use. The college sports [feed] will be my default feed if I can continue to access this through for the next 6 months---truly.''}
However, satisfaction with \System{} did not prevent participants from feeling that the cognitive load of having to formulate their intentions was too high.
For example, P13 was \textit{``very happy with the `limit posts about' when it comes to political content and news.''} Reflecting on their science-related news feed, they said \textit{``I would only see science-related news, and that made me very happy.''}
Yet, they also expressed frustration that the feeds could not learn based on their on-platform behaviors, saying \textit{``I'm very lazy and I want to be more in control, but I also want to continue to be lazy in my social media experience.''}

Participants believed that these perceptions of efforts could be partially addressed through a more integrated design.
P8 thought \textit{``it would be very convenient if you could do everything within Bluesky,''} while P1 wished for \textit{``feedback that was easy for me to give in the moment.''}
As a solution, P13 recommended a hybrid approach between intentional and inferred feeds:
\begin{quote}
\textit{``While this is a good starting point, I don't want to have to constantly edit my recommendation algorithm. I guess it'd be interesting to explore how much should be on the user versus how much should be on the recommendation algorithm going forward, and potentially a hybrid approach could be good.''} -P13
\end{quote}

To tighten the feedback loop, participants expressed a desire to iterate more with their initial natural language prompts during conversations.
P10 suggested to look at the feed and \textit{``at the same time, I can be in conversation with it''} saying that \textit{``this kind of tighter feedback loop could increase the sense of control.''}
P13 suggested that \textit{``you could just chat with [the system], and then say: 'I've been seeing too much of this lately, can you show me that?' and then it would update the prompts in the backend.''}

\subsection{Procedural transparency of intentional and personalized feeds }

Transparency was central to our design: users should see not just outputs, but how their inputs shaped the feed.  
Participants appreciated this visibility at a high level, yet struggled with the details, particularly tracing individual posts or trusting how natural language prompts were interpreted. 
These frictions reveal both the promise and the limits of procedural transparency in intentional feedbuilding.  

\subsubsection{Participants understood the feed generation process and found it easy to connect posts to preferences, but hard to trace them to their feeds precisely.}

Overall, participants could articulate how their initial prompt was transformed into sources, preferences, and eventually a ranked feed. 
For example, P7 explained their mental model of the system:

\begin{quote}
\textit{``It makes some inferences based on my first entry into the text box, and then it puts those inferences into different categories, like the include, exclude [...] and then it tried pull posts from the sources and then maybe use some sort of ranking system that follows what I want to see more and what I want to see less and also incorporating the popularity, recency, relevance in a holistic ranking system that eventually gave me this feed.''} -P7
\end{quote}

Many participants said they could ``backtrace'' what they saw into the preferences they had specified. 
As P8 put it, \textit{``I could backtrace most of what I saw to what I particularly said.''} 
P13 similarly noted that posts seemed \textit{``very directly linked to each of the descriptions,''} which made them feel more in control and insulated from irrelevant political content.

However, participants struggled to pinpoint exactly \emph{how} a given post entered their feed or which source it came from. P3 described a key limitation as \textit{``not knowing where the posts come from,''} while P12 wished the system would explicitly tell them the source of each item instead of leaving them to guess, they said:
\begin{quote}
\textit{``There's no explicit message here about which post this was in reference to. I have to infer by myself, which I can do, but I would prefer the system telling me exactly what it is. Instead of me trying to guess why a specific post was given to me...''}-P12
\end{quote}

\subsubsection{Participants questioned how natural language instructions would be translated to feed outcomes}

Some participants with prior experience working with or prompting language models reported that this knowledge shaped how they interacted with the system. P6 assumed that their description was \textit{``probably gonna be used in a prompt,''} and explained that \textit{``having some level of knowledge about how this thing would be presumably implemented''} led them to \textit{``think about how best to articulate it.''} This made the prompting step feel \textit{``less easy, because [I was wondering] what actually would be the way to phrase this?''}

Others expressed doubt about whether language models could handle complex or ambiguous requests. 
P8 felt \textit{``a bit uncomfortable giving too many instructions''} and was \textit{``not sure whether current language models are actually that good at being able to distinguish that,''} particularly when combining diverse topics. 
The participant questioned whether the system could reliably \textit{``decide what is interesting and what is not.''}
For P1, who worked in natural language processing, these concerns were even more pronounced:
\begin{quote}
\textit{``It was hard for me to write because of my work with language models. I was skeptical about what was possible and experienced a lot of hesitation in writing prompts, like, is this actually going to work?''} -P1
\end{quote}

These reflections highlight how assumptions about natural language processing can create hesitation and self-consciousness when formulating prompts, especially for technically knowledgeable users. At the same time, participants noted that this challenge might be less pronounced for more general audiences. For example, P1 also noted that \textit{``depending on the intended audience, that [previous assumptions] might not be as big a point of friction for other people.''}

\subsubsection{To trust and control feeds, participants wanted more fine-grained transparency}

Several participants emphasized that deeper visibility into the background processes would make them more effective at refining feeds. P8 noted that \textit{``if you have some insights into how it works, it can also help you really fine-tune what you're seeing.''} 
P3 similarly struggled to know which edits to make, pondering \textit{``Does this search work well? Or should I remove them? Should I boost them? I don't know, because I don't know what they're doing behind the scenes.''}
P9 sought \textit{``more information about the initial search [and how the sources performed e.g.] they have 3,000 [posts], and then they reduce it down to 2,000, and then I would like to see that kind of process. Then I will sort of understand what kind of information is actually missing out or left out.''}
In these ways, while natural language was an accessible entry point for feed constructions, participants desired greater explainability over the feeds they were investing time in.

Participants explained that increased system transparency was also needed to trust that the system was faithfully translating their instructions to their feeds. 
For example,  P8 explained:
\begin{quote}
\textit{``I think it would be super cool [...] to also have everything open source. And to be able to really understand what is happening. Also, in terms of transparency, I can trust that you don't manipulate the things that I tell [\System{}].''}-P8
\end{quote}
P9, who sees themselves as \textit{``really careful and critical using recommendation systems''} reported not trusting the system: \textit{``I don't actually know, for example, which model you're using, what kind of approach you actually designed the platform. So I'll be a little bit skeptical.''}
Without truly trusting the system, some participants were afraid of being sorted in a `filter bubble,' e.g., P9 enjoyed \System{} but still questioned \textit{``how trustable''} the system is \textit{``because of the filter bubble effect.''}

\section{Discussion}

By examining \System{}, we explore how \textit{intentional} and \textit{personalized} feedbuilding offers an alternative vision for the future of social media. 
Our findings show that participants used intentional feeds both to filter out toxic or irrelevant content and to surface new, meaningful material that was otherwise hard to find. Participants also reported feeling relieved from the pressure of having their engagement shape future content, but noted that building feeds required more effort than they were accustomed to. 
Taken together, these results outline a design space for intentional social media feeds, including fundamental tradeoffs that arise from shifting fine-grained control more directly to users.
Furthermore, they extend prior critiques of engagement-optimized feeds that document addictive or regrettable use~\cite{tranModelingEngagementDisengagementCycle2019,cho2021reflect}, demonstrating how explicit, intention-first personalization can offer benefits.

\subsection{Promoting Intentional Goal Alignment}

This work advances an \textit{intention-first} paradigm for creating and engaging with social media feeds. 
In this paradigm, systems construct feeds transparently around user goals defined directly instead of inferring preferences from engagement. 
Our findings show that participants often used \System{} in two complementary ways: to filter out toxic or irrelevant content that undermined their well-being, and to broaden their horizons by surfacing new material that aligns with their interests. This dual role distinguishes intentional feeds from engagement-based systems that can encourage compulsive scrolling and erode well-being~\cite{cho2021reflect,tran2019modeling}. 
In contrast to prior interventions, which primarily sought to reduce screen time~\cite{hinikerMyTimeDesigningEvaluating2016,koNUGUGroupbasedIntervention2015,gruningDirectingSmartphoneUse2023}, we show how intentional feeds can redirect attention toward exploration, diversification, and discovery, rather than restraint.
{We find that \System{}'s natural language interface encouraged such playfulness and nuanced preference specifications, in ways that other custom feed builders currently lack.

Our results indicate that participants felt liberated from the burden of knowing their online engagement (likes, comments, replies) could shape future feed content. 
Prior work shows that people develop ``folk theories'' to reason about algorithmic curation~\cite{devito2017algorithms,karizat2021algorithmic}. 
Our study highlights that these theories themselves can impose cognitive load, as users monitor how each click might "teach" the system and constrain their interactions. 
By decoupling intentions from engagement, \System{} relieved this burden, enabling participants to treat liking, replying, or reposting as expressive rather than instrumental. 
Our findings thus suggest that transparency and explicit intent-setting on social media feeds can help users reap more pleasure from their online experiences by encouraging freedom and spontaneity.
Today, half of U.S. teens believe that social media is detrimental to people their age: our research shows one path to making social media fun again.~\cite{pew_teens_2024}
Our work thus resonates with broader calls for more agentic, trust-building forms of transparency and control~\cite{ananny2018seeing}.

\subsection{Effort-Reward Tradeoffs}

At the same time, our study highlights that intentional feeds carry their own cognitive costs, which may be unavoidable rather than incidental.
Participants often described the act of articulating preferences as effortful, echoing findings from the personal informatics literature where users weigh the burden of self-tracking against the value of self-improvement~\cite{epstein2015lived}.
We identify that participants willing to overcome this barrier largely had strong motivations or more pragmatic expectations of effort based on prior experiences with feed builders.
With \System{}, natural language input substantially lowered the barrier to entry compared with other popular feed builders, but participants still wrestled with how best to phrase instructions or interpret the system’s responses. 
Here, procedural transparency plays a critical role, since clearer visibility into how inputs map to outputs can reduce uncertainty, reinforce users’ belief in the system’s efficacy, and make the effort of intentionality feel more worthwhile.
Our findings directly echo~\citet{vroom1964work}'s expectancy theory of motivation, which emphasizes the role of effort expectancy, instrumentation, and reward expectancy to explain what drives people to take effortful actions.
Yet, our findings also underscore a potentially fundamental tradeoff: emergent feeds, which dominate today’s platforms, offer immediacy and ease but at the expense of control~\cite{cunningham2024we,bandy2023exposure}; intentional feeds provide agency and alignment but demand greater effort, as also observed in prior systems that required active curation~\cite{bernsteinEddiInteractiveTopicbased2010,he2023cura,davisSupportingTeensIntentional2023}.
Future systems could also consider hybrid approaches that combine the ease of emergent feeds with the fidelity of intentional ones.  

\subsection{The Future of Intentional Feeds}

Intentional social media feeds will be most impactful if they are broadly and meaningfully adopted by technology companies and individuals.
This work emerged from the belief that the current engagement-first paradigm is more aligned with business incentives than human values.
However, \citet{kleinberg2024challenge} argue that there are real financial costs to over-optimizing for engagement: users may quit the platform, thereby reducing all future earnings potential.
Additionally, technology platforms increasingly face public and legal pressure to demonstrate how they prevent harmful mental health impacts~\cite{MassachusettsMetaComplaint2023, ball2025broken}.
Whether motivated by financial outcomes or public pressure, the largest technology companies have begun to tease features that better align social feeds with users’ long-term wellbeing and goals.
Since the writing of this paper, Meta’s Threads and Instagram have announced features to give users more granular control over the topics they see in their feed~\cite{threads_2025, instagram_2025}, while X’s CEO Elon Musk promised that users will be able to “adjust [their] feed dynamically just by asking Grok”~\cite{musk2025tweet}.
One plausible and even likely future for intentional feeds is therefore that they become increasingly incorporated into existing social media products. In such a future, one risk would be that these tools become a form of ``user agency theatre'' (features that lend the impression of increasing user agency but in practice have a limited impact on the outcome of interest~\citep{schneier2003beyond}), where companies use the presence of seldom-used features to maintain their current profitable and algorithmic approaches. 

At the same time, the architecture we deploy in \System{} is unlikely to be directly adopted at platform scale.
Running large LMs across planning, sourcing, and per-post curation is both computationally expensive and latency-sensitive, especially for real-time, high-throughput feeds.
In practice, any production deployment inspired by \System{} would need to dramatically reduce LM calls---for instance, by using embeddings and retrieval-style architectures to pre-compute user and item representations.
In this sense, \System{} should be understood as a pilot that probes what intent-first, LM-mediated feeds can feel like for users, and that surfaces design and engineering directions for hybrid systems in which LMs help set and refine user goals while cheaper models handle most of the online ranking workload.

\subsection{Design Implications}
\label{designimp}

Stemming from our results, we identify three key design implications for future systems that aim to support \textit{explicit} and \textit{personalized} feeds.

\subsubsection{Transparency must make control actionable}  
Participants expected high levels of procedural transparency from intentional feeds.
Because they invested effort into crafting their feeds, they wanted to understand clearly how their specifications shaped outcomes and to trace individual items back to their preferences.
This desire suggests that transparency must extend beyond exposing input configuration settings; it should also make clear how choices influence resulting feeds.  
For example, feeds could allow users to trace individual posts back to the preferences that surfaced them, preview how adjusting a filter or weight would change future content, or explain why particular items appear.  
Such mechanisms would make transparency actionable, enabling users to both understand and refine their feeds with confidence.

\subsubsection{Tightening the feedback loop}  

Participants emphasized that they wanted to express preferences and make adjustments while actively using their feeds, rather than leaving the app to modify settings elsewhere.  
Although current platforms offer limited tools to curate feed content, such as the option to ``show less like this,'' our findings suggest opportunities for richer on-the-fly feedback.
Making adjustments \textit{in situ} would reduce the cognitive load of remembering desired changes and better align edits with the moment of need.  
These findings suggest that ``teachable'' feeds, as suggested by~\citet{feng2024mapping}, could also lead to more intentional use.
In practice, enabling such features would tighten the feedback loop between configuration and consumption.  
Rather than treating feedbuilding as a separate setup activity, future systems should support lightweight, iterative changes directly within the feed interface. 
For example, users might hover over a post to trace it back to a preference, then adjust that preference on the spot, or converse with the system to refine goals in real time.

\subsubsection{Doubling down on natural language}  

Participants consistently described natural language as a flexible and intuitive way to specify feed intentions, both for initial scaffolding and later refinements.  
Compared to keyword filters or rigid configuration menus, participants found it easier to express nuanced goals in their own words.  
At the same time, participants noted opportunities to make this interaction more dynamic and conversational. 
For example, participants imagine a chatbot that could prompt them to reflect on new aspects of their intentions, or iteratively negotiate preferences through dialogue.
Others suggested a hybrid model where natural language input could be combined with lightweight edits to structured settings, giving users multiple entry points for control.  
Additionally, some feed specifications in natural language exceeded our system's predetermined capabilities for identifying sources.
The vast and creative array of intentions that participants wrote suggests that agentic AI capabilities may be well deployed for helping users align feeds to varying intentions.
Taken together, these findings highlight natural language not only as a low-effort entry point, but also as a foundation for richer, more imaginative, and more adaptive forms of intentional feedbuilding.

\subsection{Limitations}  

Our study has several limitations that shape the interpretation of our findings.  
First, because most mainstream social media platforms are closed-source and not extensible via APIs/protocols, we implemented \System{} on Bluesky, a decentralized service built on the AT Protocol.  
While this choice enabled rapid prototyping and external feed customization, it also constrained the evaluation.  
Bluesky currently hosts less content and weaker engagement-based feeds compared to platforms like Twitter or Instagram, meaning participants’ experiences with \System{} may not fully generalize to larger, more mature ecosystems.  

Second, our recruitment strategy was necessarily limited to Bluesky users, who represent a distinct slice of the population.  
Prior work suggests this community is skewed in both political alignment and social media practices~\cite{nogara2025longitudinalanalysismisinformationpolarization}.  
Our participant pool also leaned toward academics, further narrowing the diversity of perspectives included in the study.  

Third, participants interacted with \System{} for approximately 11 days.  
Although this duration was sufficient to surface short-term impressions and design needs, longer-term adoption might reveal different patterns, including sustained benefits or emergent challenges that only arise with habitual use.  

Fourth, while \System{} demonstrates the potential of decentralized protocols like Bluesky's AT Protocol to enable external feed customization, current architectural constraints prevent developers from embedding customization features directly into the client.  
Comparable ecosystems, such as Farcaster,\footnote{\url{https://docs.farcaster.xyz/}} already support richer forms of intervention by allowing embedded mini-applications within posts.  
Extending protocol openness to user-experience layers would enable tighter, more expressive feedback loops and make intentional feedbuilding feel seamless.

Fifth, \System{} was built and optimized for content-based filtering. Users sometimes found new use cases we had not anticipated, such as asking \System{} to fetch and analyze their existing network. In such unanticipated scenarios, system restrictions could cause friction for users, who may not have known their request could not be translated to a feed outcome. This system constraint partly motivates this paper's emphasis on procedural transparency.

Finally, intentional feeds themselves may carry risks.  
A few participants worried that greater control could inadvertently reinforce echo chambers, filtering out not only unwanted noise but also valuable opposing perspectives.  
Future work should examine these downstream risks and investigate design strategies that balance user agency with exposure diversity.

\section{Conclusion}  

In this work, we investigated an alternative to engagement-based social media feeds that prioritizes user attention over long-term well-being.  
We proposed an approach that centers \textit{explicit} and \textit{personalized} feeds, enabling users to articulate their goals directly in natural language and transparently shape the algorithms that serve them.  
\System{} implements this approach through a platform-agnostic framework with four stages: planning, sourcing, curating, and ranking.  
Through a multi-week field study with fifteen participants, we found that users employed \System{} both to filter out toxic or irrelevant content and to broaden their horizons by discovering new material.  
Participants described feeling liberated from the burden of curating their behavior to “teach” opaque algorithms, and valued the sense of agency afforded by transparent, intention-first design.  
At the same time, they noted the cognitive effort involved in configuring explicit feeds and emphasized the importance of actionable transparency, tighter feedback loops, and conversational natural language interfaces.  
Overall, our findings demonstrate that intentional feedbuilding is both feasible and desirable, offering a concrete path toward social media that better aligns with user goals and values.  

\begin{acks}
We thank the Hasler Foundation for partial support of OEM during his master's thesis, and the Cornell Tech Social Technologies Lab for feedback on earlier versions of this work.
\end{acks}

\bibliographystyle{ACM-Reference-Format}
\bibliography{main}

\appendix

\section{Additional User Interface Screenshots}
We show additional screenshots of \System{}'s user interface on Figure ~\ref{fig:system-overview2} and Figure ~\ref{fig:ranking-presets}.

\section{Database construction}
\label{app:db-construction}
To construct the database of feeds, starter packs, and lists, we scraped 'blueskydirectory.com' and 'bskyinfo.com' using BeautifulSoup to extract feeds, lists, and starter packs for addition to our database. We scraped feeds that had at least two likes. We then stored the database in the server and made it accessible via an internal API that takes search queries and returns all entries that match the query in their description or title. The database used for our field study had 1277 feeds, 205 starter packs, and 300 lists.

\begin{figure*}[p]
    \centering
    
    \begin{minipage}[t]{0.45\textwidth}
        \centering
        \raisebox{-0.5\height}{\includegraphics[width=0.7\textwidth]{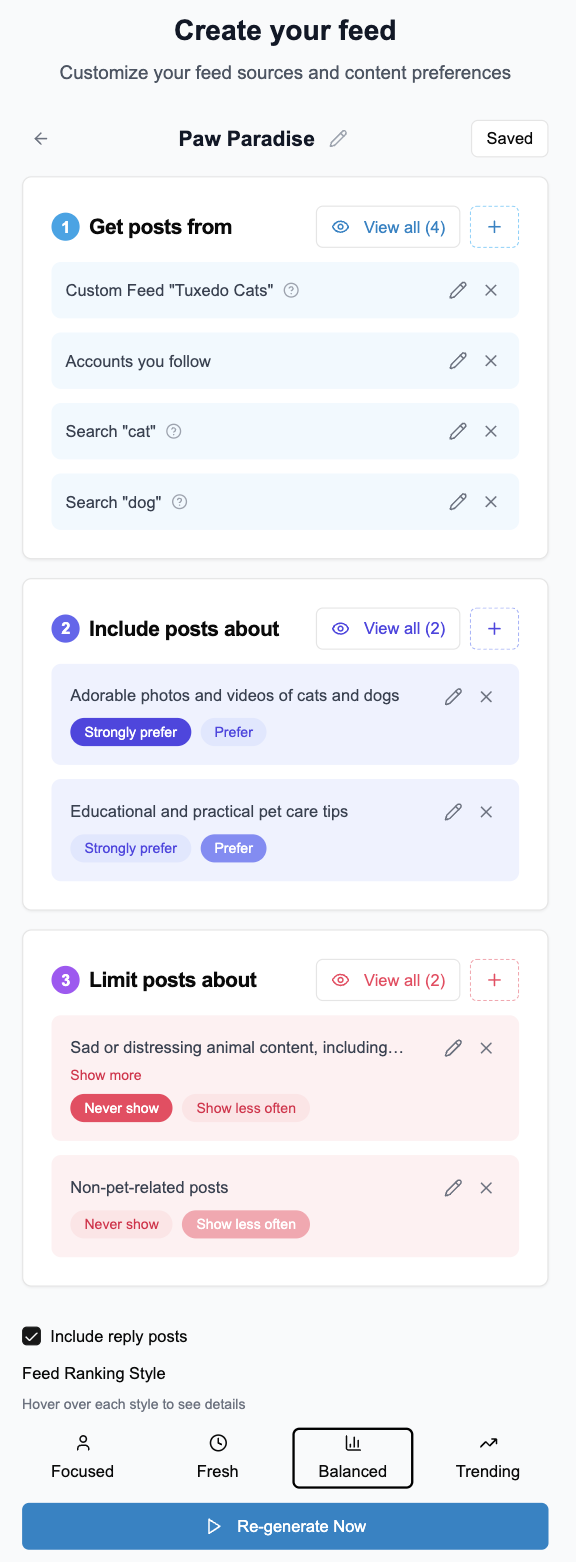}}
        
        \vspace{0.3cm}  
        
        \textbf{(a)} Feed configuration page \\
        \footnotesize{Where users configure their sources, preferences and generate their feeds.}
    \end{minipage}
    \hspace{1cm}
    \begin{minipage}[t]{0.45\textwidth}
        \centering
        \raisebox{-0.5\height}{\includegraphics[width=\textwidth]{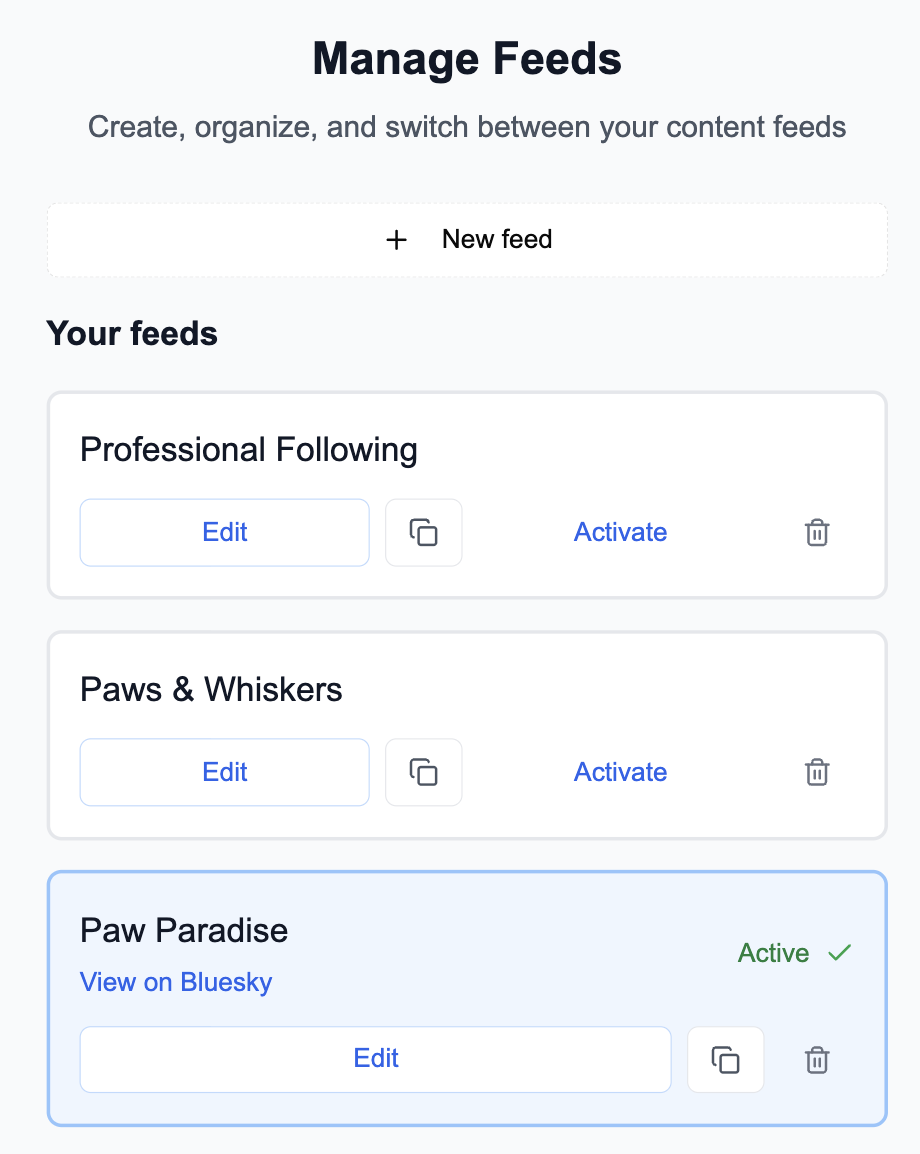}}
        
        \vspace{0.3cm}  
        
        \textbf{(b)} Manage Feeds page \\
        \footnotesize{Where users can activate, duplicate and delete feeds from their accounts.}
    \end{minipage}
    
    \caption{Feed management interfaces in \System{}. (a) Feed configuration page, where users customize their sources, set inclusion and exclusion prompts, and select ranking styles. (b) Manage feeds page, where users can view, activate, duplicate, and edit multiple feeds from a single account.}
    \label{fig:system-overview2}
\end{figure*}

\begin{figure*}[p]
  \centering

  \begin{subfigure}{0.47\textwidth}
    \centering
    \includegraphics[width=\linewidth]{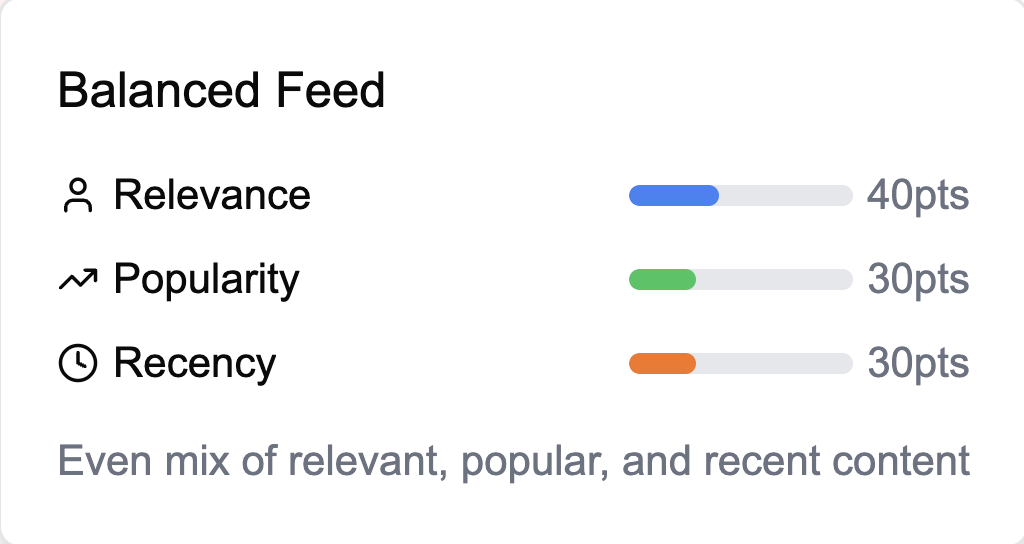}
    \label{fig:tl}
  \end{subfigure}\hfill
  \begin{subfigure}{0.47\textwidth}
    \centering
    \includegraphics[width=\linewidth]{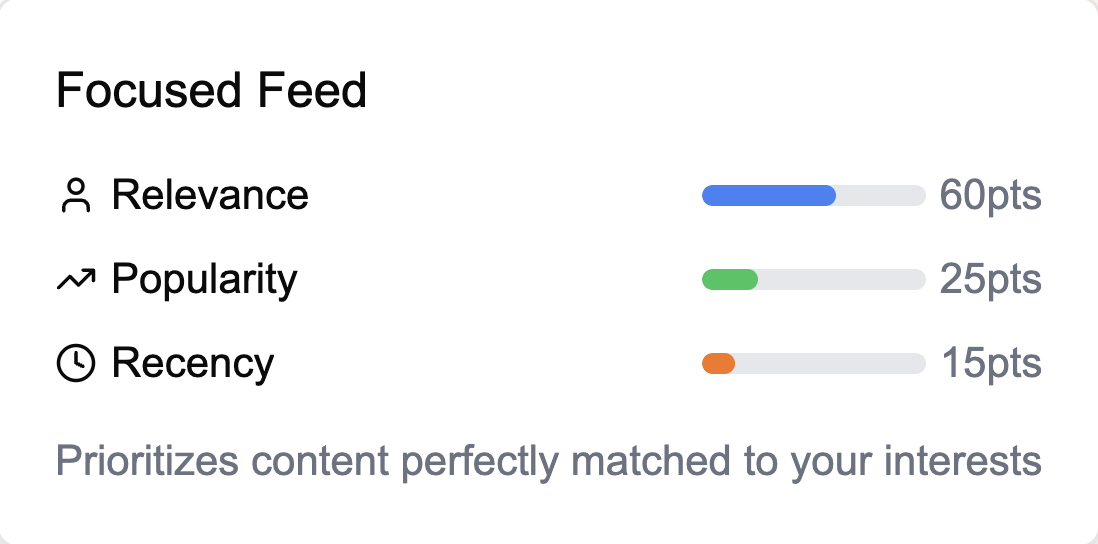}
    \label{fig:tr}
  \end{subfigure}

  \vspace{0.5em}

  \begin{subfigure}{0.45\textwidth}
    \centering
    \includegraphics[width=\linewidth]{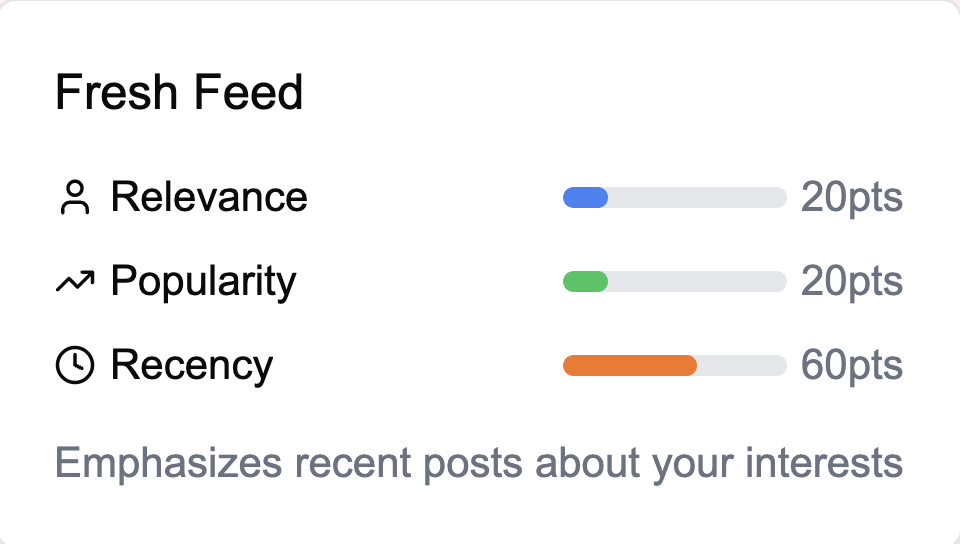}
    \label{fig:bl}
  \end{subfigure}\hfill
  \begin{subfigure}{0.48\textwidth}
    \centering
    \includegraphics[width=\linewidth]{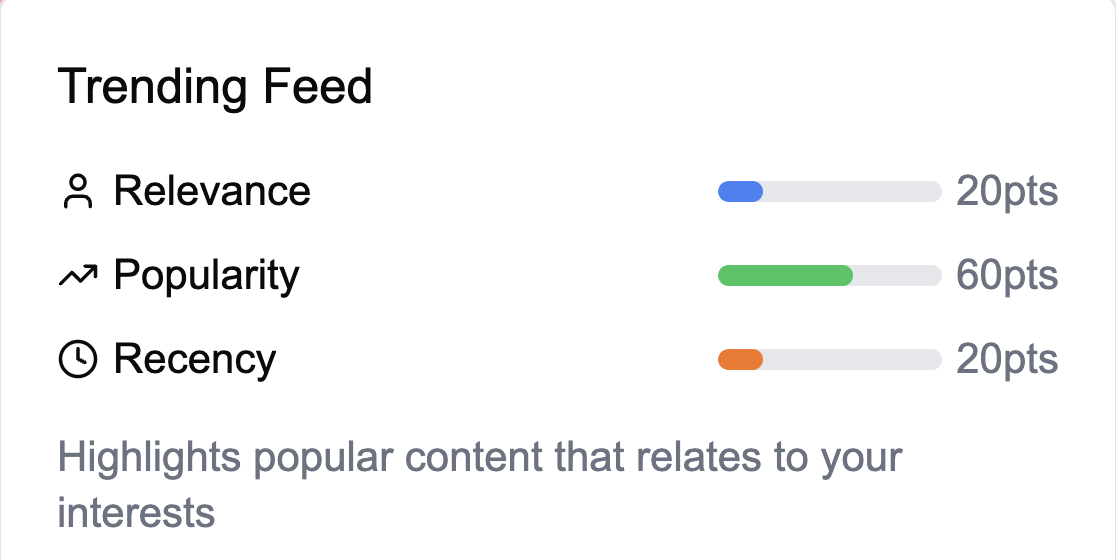}
    \label{fig:br}
  \end{subfigure}

  \caption{Ranking presets in \System{}. Users can choose between four styles: Balanced (equal emphasis on relevance, popularity, and recency), Focused (prioritizes relevance), Fresh (emphasizes recency), and Trending (emphasizes popularity). Each preset adjusts point weights that contribute to the final ranking algorithm.}

  \Description{This figure shows the four ranking presets that Bonsai offers to let users control how their feeds are ordered. Each preset adjusts the relative weights assigned to three criteria: relevance, popularity, and recency. The ``Balanced Feed'' splits weights fairly evenly, with a slight emphasis on relevance, producing an even mix of relevant, popular, and recent content. The ``Focused Feed'' heavily favors relevance, giving less weight to popularity and recency, so that content closely aligned with the user’s stated goals is prioritized. The ``Fresh Feed'' emphasizes recency, surfacing the newest posts while still considering relevance and popularity to a smaller degree. Finally, the ``Trending Feed'' highlights popularity, prioritizing posts that are widely engaged with, while still incorporating recency and relevance. Each point in the weight system contributes incrementally to the ranking formula, allowing these presets to serve as simple yet powerful shortcuts for tailoring how Bonsai assembles a user’s feed.}
  \label{fig:ranking-presets}

  \Description{This figure presents two core user interfaces from  Bonsai that support feed customization and management. On the left, the feed configuration page illustrates how users create a custom feed, here named ``Paw Paradise.'' The interface allows users to select sources such as accounts, hashtags, or search terms, specify inclusion rules like always showing posts with cats and dogs, and set exclusion rules such as hiding distressing or promotional content. At the bottom, users can also choose a ranking style (Focused, Fresh, Balanced, or Trending) depending on whether they want to prioritize relevance, recency, or popularity. On the right, the Manage Feeds page shows how users can oversee multiple feeds at once, with examples like "Professional Following" and "Paws \& Whiskers." From this page, users can edit, activate, duplicate, or delete feeds, and view their active feed directly on Bluesky. Together, these pages demonstrate how Bonsai combines fine-grained configuration with simple management tools for intentional, personalized feed building.}
\end{figure*}

\clearpage

\section{Ranking Algorithm}
\label{app:borda}
The Weighted Borda Count is a rank aggregation method that combines multiple ranked lists into a single ordering.
Each post receives points based on its position within each ranking dimension (e.g., relevance, recency, popularity).
Higher ranks receive more points, and these points are multiplied by user-specified weights that reflect the importance of each dimension.
The final score for each post is the weighted sum of its points across all dimensions, and posts are ordered accordingly.
Below, we reproduce the pseudocode for the algorithm used.

\begin{algorithm}[H]
\caption{Weighted Borda Count for Social Media Feed Ranking}
\label{alg:weighted-borda-count}
\begin{algorithmic}[1]
\REQUIRE Posts $P = \{p_1, p_2, \ldots, p_n\}$, user preferences $U$, weights $W = \{w_r, w_p, w_c\}$ where $w_r + w_p + w_c = 1$
\ENSURE Ranked list of posts $R$

\STATE \textbf{Phase 1: Generate Individual Rankings}
\STATE $R_{\text{relevance}} \gets []$
\FOR{each post $p \in P$}
    \STATE $(include, priority) \gets \text{LLM-Evaluate}(p, U)$ 
    \IF{$include = \text{True}$}
        \STATE $R_{\text{relevance}}.\text{append}((p, priority))$
    \ENDIF
\ENDFOR
\STATE Sort $R_{\text{relevance}}$ by $priority$ descending, assign ranks 1 to $|R_{\text{relevance}}|$

\STATE $R_{\text{recency}} \gets []$
\FOR{each post $p \in P$}
    \STATE $t \gets p.\text{created\_at}$ 
    \STATE $R_{\text{recency}}.\text{append}((p, t))$
\ENDFOR
\STATE Sort $R_{\text{recency}}$ by $t$ descending, assign ranks 1 to $|P|$

\STATE $R_{\text{engagement}} \gets []$
\FOR{each post $p \in P$}
    \STATE $e(p) \gets p.\text{likes} + 3 \times p.\text{reposts} + 2 \times p.\text{replies}$ 
    \STATE $R_{\text{engagement}}.\text{append}((p, e(p)))$
\ENDFOR
\STATE Sort $R_{\text{engagement}}$ by $e(p)$ descending, assign ranks 1 to $|P|$

\STATE \textbf{Phase 2: Weighted Borda Aggregation}
\STATE $P_{\text{eligible}} \gets \{p : p \in R_{\text{relevance}}\}$ 
\STATE $\text{final\_scores} \gets []$
\FOR{each post $p \in P_{\text{eligible}}$}
    \STATE $r_r(p) \gets \text{rank of } p \text{ in } R_{\text{relevance}}$
    \STATE $r_c(p) \gets \text{rank of } p \text{ in } R_{\text{recency}}$
    \STATE $r_e(p) \gets \text{rank of } p \text{ in } R_{\text{engagement}}$
    \STATE $s(p) \gets w_r \cdot r_r(p) + w_p \cdot r_e(p) + w_c \cdot r_c(p)$ 
    \STATE $\text{final\_scores}.\text{append}((p, s(p)))$
\ENDFOR

\STATE Sort $\text{final\_scores}$ by score ascending 
\STATE \textbf{return} $[p \text{ for } (p, s) \text{ in } \text{final\_scores}]$
\end{algorithmic}
\end{algorithm}

\section{Co-Occurrence of User Intents in Their Natural Language Instructions}

\begin{figure}[H]
    \centering
    \includegraphics[width=\linewidth]{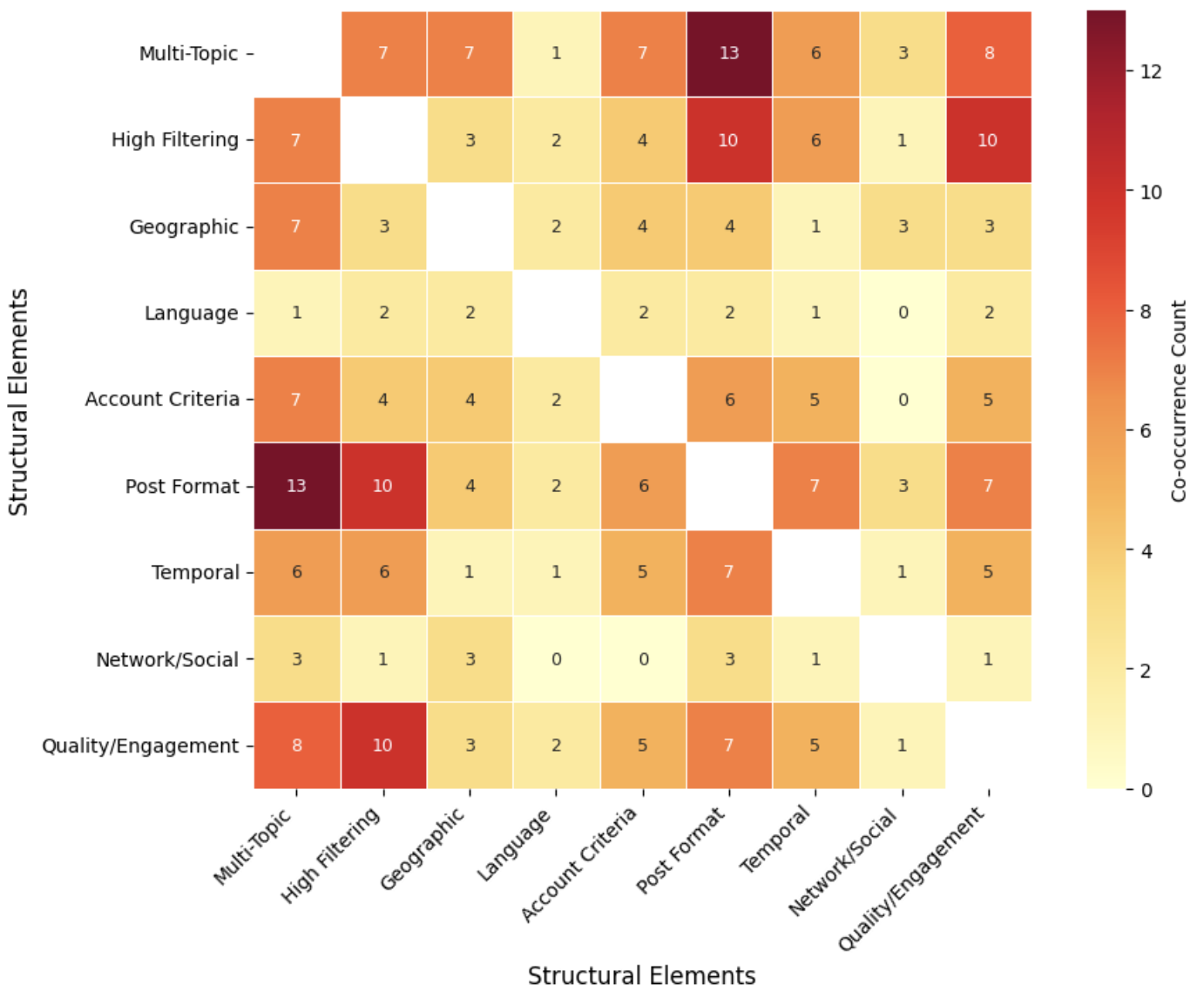}
    \caption{Co-occurrence heatmap of structural elements in user prompts. Rows and columns represent categories such as Multi-topic, High Filtering, Geographic, etc. Darker colors indicate higher co-occurrence counts, showing that participants often combined multiple structural elements (e.g., Geographic and Post format) when describing their feed preferences.}
\label{fig:cooccurrences}    

\Description{This figure presents a heatmap showing how different structural elements co-occurred in participants’ natural language feed instructions. The vertical and horizontal axes both list categories such as ``Multi-topic,'' ``High Filtering,'' ``Geographic,'' ``Language,'' ``Account Criteria,'' ``Post Format,'' ``Temporal,'' ``Network/Social,'' and ``Quality/Engagement.'' Each cell shows the number of user prompts that contained both features, with darker red colors indicating higher counts and lighter yellow colors indicating lower counts. For example, ``Post Format'' and ``High Filtering'' co-occurred frequently, as did ``Account Criteria'' with ``Multi-topic.'' In contrast, some combinations, like ``Network/Social'' and ``Language,'' appeared rarely. This visualization highlights patterns in how users structured their feed intentions, showing that people often combined multiple dimensions of specification (e.g., filtering and post type) when describing what they wanted from their personalized feeds.}
\end{figure}

\section{Interview Protocol (1st Interview)}\label{interview-protocol-1}

We include a simplified interview protocol for the both our interviews.  The interviews were semi-structured; not all questions were consistently asked to participants, and interviewers had the freedom to ask other questions.

\subsection{Part 1: Social Media Experience}

\subsubsection{Platforms used}

\begin{enumerate}
    \item What social media platforms do you currently use?
    
    \item Platform-by-Platform Deep Dive \textit{(choose the main 1 platform they use + Bluesky)}
    \begin{itemize}
        \item Let's talk about [Platform X]. Why do you use this platform and what do you get out of using it?
        \item How long have you been/How much time do you spend on the platform? Which feeds do you use?
        \item How do you feel about the control you have over what you see on [Platform X]?
        \item Do you have a sense of how the feeds are generated? How do you think [Platform X] decides what to show you in your feed?
    \end{itemize}
\end{enumerate}
    
\subsubsection{Overall Patterns}

    \begin{enumerate}
        \item Now that we've talked about each platform specifically, what are your overall impressions of how you use social media?
        \item If you could have more control over what you see on social media, what would that look like?
    \end{enumerate}

\subsection{Part 2: Feed Creation Think-Aloud Walkthrough}

\subsubsection{Setup}
\begin{enumerate}
    \item Ask to share screen
    \item Send them the link to the app
    \item Ask to use an app password
\end{enumerate}
\subsubsection{Think-Aloud Protocol}
\begin{enumerate}
    \item Now I'm going to show you a tool that lets you create your own social media feed using natural language. I'd like you to think out loud as you use it - tell me what you're thinking, what you're trying to do, what's confusing, what makes sense. Just say whatever comes to mind.
    \item When they go quiet: ``What are you thinking right now?''
    \item When they seem stuck: ``Talk me through what you're trying to do''
    \item When they make decisions: ``What made you choose that?''
    \item Whenever they mention something, go deeper
\end{enumerate}

\subsubsection{Reflection}

\begin{enumerate}
    \item How was that experience for you?
    \item How does the feed look like in comparison to what you expected?
    \item How does creating your own feed compare to using [their main platform]?
    \item How did the interface feel like?
    \item What is your impression of how the feed is being generated?
    \item In what ways did this experience make you feel like you had control over the feed, or make you feel a lack of control?
    \item How do you imagine a feed based on your expressed natural language might diverge from a regular algorithmic feed?
    \item Do you think you are as likely to use this kind of feed as an algorithmic feed?
    \item What kinds of advantages can you see from this approach? What disadvantages?
\end{enumerate}

\section{Interview Protocol (Follow-up interview)}\label{interview-protocol-2}

\label{app:follow-up-interview-guide}

\subsection{Overall Experience}
\begin{enumerate}
    \item How would you describe your overall experience using \System{} this past week?
    
    \item How often did you find yourself using your custom feed versus your usual bsky feed/other platforms?
\end{enumerate}

\subsection{Quick Walkthrough}
\begin{itemize}
    \item I'd like you to please share your screen on our app and show me how you used the system this week briefly.
    \item Did you create multiple feeds for different purposes, or stick with one?
    \item If multiple: How did you decide what to separate? If single: Why one feed?
    \item Explain to me your motivation for creating each feed.
\end{itemize}

\subsection{Follow-up questions}

\begin{enumerate}
    \item Thinking about the content you saw in your custom feed, how did it compare to what you typically see? Did you discover new content? Was it relevant? Quality differences?
    \item Can you give me an example of something you saw in your custom feed that you wouldn't have seen elsewhere? How did that make you feel?
    \item How much control did you feel you had over your feed? What specific aspects made you feel more/less in control?
    
    \item Were there moments when you felt like your changes had a direct impact on what you saw in the feed?
    \item Let's talk about describing your feed preferences in natural language/plain english. How did that approach feel as a way to communicate your intentions to the system? Did you feel like the system understood what you were asking for when you wrote your description? Any examples where it got it right or wrong? What made the natural language approach work well for you, and what didn't?

    \item How did the experience of intentionally designing your feed compare to adapting to an algorithmic feed/using other curation methods (blocking, following, muting keywords)?
    
    \item Did you feel like you sacrificed something to obtain these benefits?
    
    
    \item Did the intentional feed design process change how you think about what you want from social media?
    
    
    \item How well could you understand why certain content was appearing in your feed?
    \item When you made changes, how clear was it what effect those changes would have?
    \item{You had access to different levels of control - from high-level descriptions down to specific source management. To what extent did you attempt to use both and how did having these different levels of control affect your experience?}
\end{enumerate}

\subsection{Wrap-up and Meta Reflection}
\begin{enumerate}
    \item How likely would you be to continue using this type of feed system if it were available permanently and reliable?    
    \item If not, what would make you more likely to adopt it?
    \item Looking back on both interviews, has your perspective on social media feeds or content control changed at all?
    \item Is there anything significant about your experience that we haven't covered?
\end{enumerate}

\section{Prompts for Scoring}
\label{app:prompts}

We also provide the prompt used below for the curation step of the pipeline to evaluate a post's relevance. This score, combined with the engagement and recency scores (described exactly in the algorithm), is combined through the Weighted Borda Count algorithm to obtain the final ranking.

\begin{framed}
\begin{Verbatim}[fontsize=\fontsize{6}{8}]
EVALUATION GUIDELINES:
1. NEVER SHOW items should be completely blocked - if the post relates to 
these topics at all, 
exclude it
2. SHOW LESS OFTEN items should be filtered out unless they're
particularly relevant or 
high-quality
3. SHOW MORE OFTEN items should be given preference - include these posts
even if they're 
borderline relevant
4. ALWAYS SHOW items should be prioritized - include these posts if they 
relate to the topic at all
5. NEUTRAL CONTENT (doesn't match preferences but not excluded) should be
INCLUDED LIBERALLY for feed diversity

CONTENT MATCHING:
- Consider the main topic, theme, and context of the post
- Pay attention to hashtags, mentions, and embedded content
- Look for indirect references and related concepts
- Be more strict with exclusions (NEVER/LESS) and more lenient with 
preferences (MORE/ALWAYS)
- BE INCLUSIVE of neutral content that doesn't match
preferences but isn't excluded

PRIORITY SCORING (for included posts):
- 10: Perfect match for ALWAYS SHOW preferences (exact topic, high 
quality)
- 8-9: Good match for ALWAYS SHOW preferences (related topic, good 
quality)
- 6-7: Perfect match for SHOW MORE OFTEN preferences (exact topic)
- 4-5: Good match for SHOW MORE OFTEN preferences (related topic)
- 3-4: Neutral/general interest content (no strong preference match) - 
INCLUDE THESE FOR DIVERSITY
- 2-3: Borderline relevant but acceptable content - 
INCLUDE THESE FOR DISCOVERY
- 1: Barely relevant but not harmful content - INCLUDE THESE FOR VARIETY

DECISION PROCESS:
1. First check if the post matches any NEVER SHOW exclusions - if yes, 
exclude immediately
2. Then check if it matches SHOW LESS OFTEN exclusions - if yes, be very 
strict about including it
3. Check if it matches any ALWAYS SHOW preferences - if yes, include it 
with priority 8-10
4. Check if it matches SHOW MORE OFTEN preferences - if yes, include it 
with priority 4-7
5. If no strong matches AND no exclusions match, DEFAULT TO INCLUDE with 
priority 2-4 for content diversity
6. Only exclude if the content is clearly harmful, spam, or matches 
exclusions

CONTENT DIVERSITY PRINCIPLE:
- The user wants a diverse feed that includes content beyond their 
specific preferences
- If content doesn't match exclusions, lean toward including it for
discovery and variety
- Neutral content helps prevent echo chambers and exposes users to new 
topics
- Only exclude content that is clearly unwanted, not just unrelated to
preferences

RESPONSE FORMAT:
- If excluding: respond with 'no'
- If including: respond with 'yes:PRIORITY' where PRIORITY is a 
number 1-10
- Examples: 'yes:10', 'yes:7', 'yes:3', 'no'
- When in doubt about neutral content, choose 'yes:3' rather than 'no'

'yes:X' = include this post in the user's feed with priority X
'no' = exclude this post from the user's feed
\end{Verbatim}
\end{framed}

\section{Planning Agent Details}
\label{app:agent}

In the planning phase, an LLM agent helps identify relevant feeds, starter packs, and lists for a user query. The agent operates in an action–observation loop: it first generates keywords from the user prompt, then queries a database via keyword matching over titles and descriptions. After proposing an initial set of sources, the user refines their preferences through the feed configuration interface. The agent then reruns the same keyword-generation and retrieval process using the updated preferences, and the user reviews and selectively includes additional suggested sources before proceeding to feed generation.
We provide the prompt provided to the agent to generate keywords below:

\begin{framed}
\begin{Verbatim}[fontsize=\fontsize{6}{8}]
Based on these preferences, generate 5-8 SHORT SINGLE-WORD search
terms that would help find relevant feeds, starter packs, and 
lists in a content discovery database.

CRITICAL: Always start with the most OBVIOUS, BASIC terms from 
their preferences:
- If they mention "dogs and cats" → include "dogs", "cats", "pets"
- If they mention "machine learning" → include "machine", 
"learning", "ai", "ml"
- If they mention "Love Island US" → include "love", "island", "reality"
- If they mention "cooking" → include "cooking","recipes", "food"

Then add 2-3 related terms for broader discovery.

Return a JSON object with:
- "search_terms": array of 5-8 single words starting with the
most obvious terms
- "reasoning": brief explanation of why these terms were chosen

Example for "I want posts about cute dogs and cats":
{{
  "search_terms": ["dogs", "cats", "pets",   "animals", "cute",
  "adorable"],
  "reasoning": "Start with obvious terms   dogs/cats/pets, then
  add related animal and emotion terms"
}}

IMPORTANT: 
- ALWAYS include the most obvious, basic words from their preferences
FIRST- Don't overcomplicate - use simple, direct terms
- Use only single words for better database matching
\end{Verbatim}
\end{framed}

\section{Generative AI Disclosure}
Generative AI tools were used to enhance the search for related works and refine the writing and formatting of this manuscript.
We followed guidance from~\citet{schroeder2025large}, who provide recommendations to identify legitimate uses of AI for research efficiency while safeguarding the qualitative sensemaking process.
Specifically, Claude, ChatGPT, and Elicit were used to find relevant research papers for both the related works and discussion sections (alongside non-Generative AI tools, like Google Scholar and existing Zotero libraries).
After the Discussion had been written, ChatGPT was used to refine and streamline the content, and then manually edited again by the authors.
Claude was also used for specific formatting tasks, such as generating table formats or translating the interview protocol to LaTeX.
Where generative AI has been used for editing, the authors certify that they have read, adapted, and corrected the text as necessary, and stand behind the resulting work.

\end{document}